\documentclass[sn-basic]{sn-jnl}
\usepackage{graphicx}%
\usepackage{hhline, multirow}%
\usepackage{amsmath,amssymb,amsfonts}%
\usepackage{mathrsfs}%
\usepackage[title]{appendix}%
\usepackage{xcolor}%
\usepackage{textcomp}%
\usepackage{manyfoot}%
\usepackage{booktabs}%
\usepackage{algorithm}%

\usepackage{algorithmicx}%
\usepackage{algpseudocode}%
\usepackage{listings}%
\usepackage{array}%
\usepackage{caption}
\usepackage{lscape}
\usepackage{adjustbox}
\usepackage{tabularx}
\usepackage{adjustbox}%
\usepackage{subcaption} 
\usepackage{booktabs} 
\usepackage{hyperref} 
\newcolumntype{P}[1]{>{\centering\arraybackslash}p{#1}}
\usepackage{orcidlink}

\begin{document}

\title{Explainable AI: Comparative Analysis of Normal and Dilated ResNet Models for Fundus Disease Classification}


\author*[1]{\fnm{P.N.} \sur{Karthikayan} \orcidlink{0000-0003-2508-3319}}\email{coe20d002@iiitdm.ac.in}

\author[1]{\fnm{Yoga Sri Varshan} \sur{V} \orcidlink{0009-0007-8440-8060}}\email{ced18i058@iiitdm.ac.in}

\author[1]{\fnm{Hitesh Gupta} \sur{Kattamuri} \orcidlink{0009-0000-2751-1802}}\email{cs20b1127 @iiitdm.ac.in}

\author[1]{\fnm{Umarani} \sur{Jayaraman} \orcidlink{0000-0002-9676-6291}}\email{umarani@iiitdm.ac.in}

\affil*[1]{\orgdiv{Department of Computer Science and Engineering}, \orgname{Indian Institute of Information Technology Design and Manufacturing, Kancheepuram}, \orgaddress{\city{Chennai}, \postcode{600127}, \state{Tamil Nadu}, \country{India}}}


\abstract{This paper presents dilated Residual Network (ResNet)  models for disease classification from retinal fundus images. Dilated convolution filters are used to replace normal convolution filters in the higher layers of the ResNet model (dilated ResNet) in order to improve the receptive field compared to the normal ResNet model for disease classification. 
This study introduces computer-assisted diagnostic tools that employ deep learning, enhanced with explainable AI techniques. These techniques aim to make the tool's decision-making process transparent, thereby enabling medical professionals to understand and trust the AI's diagnostic decision. They are particularly relevant in today's healthcare landscape, where there is a growing demand for transparency in AI applications to ensure their reliability and ethical use.
The dilated ResNet is used as a replacement for the normal ResNet to enhance the classification accuracy of retinal eye diseases and reduce the required computing time.
The dataset used in this work is the Ocular Disease Intelligent Recognition (ODIR) dataset which is a structured ophthalmic database with eight classes covering most of the common retinal eye diseases. The evaluation metrics used in this work include precision, recall, accuracy, and F1 score. In this work, a comparative study has been made between normal ResNet models and dilated ResNet models on five variants namely ResNet-18, ResNet-34, ResNet-50, ResNet-101, and ResNet-152. The dilated ResNet model shows promising results as compared to normal ResNet with an average F1 score of 0.71, 0.70, 0.69, 0.67, and 0.70 respectively for the above respective variants in ODIR multiclass disease classification..}

\keywords{Normal ResNet, Dilated ResNet, Explainable AI, Fundus Images, Disease Classification, ODIR Dataset, Eye Disease}



\maketitle

\section{Introduction}
\begin{figure}[htbp]
\centerline{\includegraphics[scale=0.6]{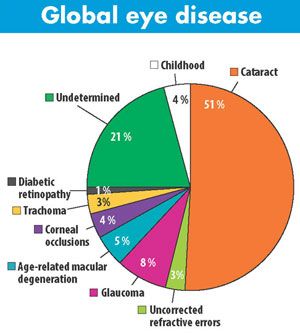}}
\caption{Source: Global Health Matters, 2012 \cite{GHM2012}}
\label{GHM}
\end{figure}

In 2022, according to World Health Organization (WHO) globally, at least 2.2 billion people have a near or distance vision impairment. 
These people include those with moderate or severe distance vision impairment or blindness due to unaddressed refractive error (88.4 million), cataract (94 million), age-related macular degeneration (8 million), glaucoma (7.7 million), diabetic retinopathy (3.9 million), as well as near vision impairment caused by unaddressed presbyopia (826 million) \cite{WHO}. The percentage of global eye disease as reported in Global Health Matters \cite{GHM2012} is shown in Figure \ref{GHM}.

The leading causes of vision impairment are i) cataract
ii) age-related macular degeneration iii) diabetic retinopathy iv) glaucoma v) uncorrected refractive errors. There is substantial variation in the causes between and within countries according to the availability of eye care services, their affordability, and the eye care literacy of the population \cite{WHO}.

\begin{figure}[htbp]
\centerline{\includegraphics[scale=0.8]{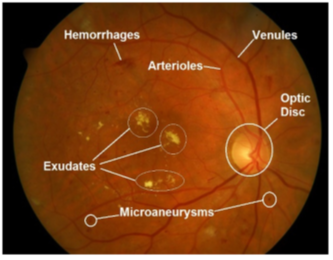}}
\caption{Retinal fundus image}
\label{fundus}
\end{figure}

Fundus images provide details about a patient's retina and are captured with a specialized fundus camera. The retinal macula, optic disc, fovea, and blood vessels constitute some of the features noticed in the image. Because of these features, the fundus imaging method is more cost-effective and better suited for non-invasive screening. Fundus photography is a screening tool used by qualified medical professionals such as optometrists, ophthalmologists, orthoptists, and others to monitor the progression of certain eye diseases. 
Few possible eye diseases that can be seen from retinal fundus images are listed below and the same is highlighted in Figure \ref{fundus}.

\begin{itemize}
\item Cataract \cite{askarian2021detecting} - eye's vitreous opacity, which is caused by protein denaturation, resulting in the blurring of the basic fundus structures

\item Glaucoma \cite{manassakorn2022glaunet} - the optic disc and optic cup in the retinal (glaucoma) images are significantly enlarged compared to normal eyes.

\item  Diabetic retinopathy \cite{article} fundus images have abnormalities like microaneurysms, soft exudates, hard exudates, and hemorrhages.
\end{itemize}

\subsection{Need for Automatic Disease Classification}
Analyzing fundus images is an exhausting and tedious task even for trained experts as these images comprise complex components such as optic disc and cup, blood vessels, macula, and so on. The vascular structure of the retina is so complex and diseases that affect these structures are so subtle. As a result, it is difficult to identify these small changes by human experts, which makes the task of manual disease classification so difficult. Also, manual disease classification is laborious and time-intensive, vulnerable to inter-rater variability, and has reduced efficiency.
Furthermore, the increasing number of patient data adds to the challenge of clinical routines such as diagnosis, treatment, and monitoring. As a result, automated approaches for disease classification from retinal images can be helpful in clinical settings.
Also, it can help in identifying vessel health and early risk prediction of diseases like stroke and heart attacks.

Automatic disease classification of retinal fundus images is crucial for early detection and diagnosis of conditions like cataract, diabetic retinopathy, age-related macular degeneration, and glaucoma. 
Healthcare professionals can effectively screen numerous images by automating the classification process. This ensures prompt intervention and may help prevent blindness or vision loss. Furthermore, automation lowers human error and inter observer variability while improving consistency and standardization in interpretation. By allowing specialists to concentrate on cases that need more investigation, this strategy maximizes the use of healthcare resources. It also makes remote monitoring and telemedicine applications possible for patients in underprivileged areas.
Overall, automatic disease classification of retinal fundus images holds the potential to transform the diagnosis, treatment, and research of eye diseases, resulting in better patient outcomes and more effective healthcare.


\subsection{ Gaps in Research}
Deep convolution neural network models have been used extensively for image classification tasks. The ResNet model is one such model that is widely used. Having said that, normal ResNet models, while effective for a wide range of computer vision tasks, may not be the most suitable choice for disease classification from retinal fundus images due to several reasons:

\textbf{Restricted Receptive Field}: Normal convolutional layers, which are commonly used in ResNet models, have a limited receptive field. In the context of retinal images, where diseases may manifest in subtle or localized features across the image, models with a restricted receptive field may find it difficult to extract important contextual information required for precise disease classification.

\textbf{Spatial Hierarchies and Context}: Retinal images contain complex structures at multiple spatial scales, including blood vessels, optic disc, and macula. A normal ResNet model may not effectively capture these hierarchical spatial relationships and contextual cues, leading to poor performance in distinguishing between different retinal diseases.

\textbf{Loss of Spatial Resolution}: Down sampling in deep ResNet models often result in small output feature maps at the end of the network, which may lead to a loss of spatial resolution. Preserving fine-grained spatial details in retinal images is crucial in detecting subtle changes related to disease.

Overall, while ResNet models have demonstrated success in various computer vision tasks, their inherent limitations in capturing spatial hierarchies, preserving spatial resolution, and handling complex data distributions make them less suitable for disease classification from retinal fundus images. Instead, models such as dilated ResNet, which address these challenges by incorporating dilated convolutions are better suited for this specific application domain.

\subsection{Contributions of the work}

The important contributions of this work are highlighted below,
\begin{itemize}
\item A comparative study between normal ResNet models and dilated ResNet models for disease classification on the ODIR dataset has been conducted.
\item Five different ResNet variants of varying network depth have been used in the study, namely ResNet-18, ResNet-34, ResNet-50, ResNet-101 and ResNet-152.
\item The accuracy and F1 score results obtained in dilated ResNet models outperform normal ResNet models. It has been observed that the average accuracy has improved from 0.70 to 0.79 while the F1 score has improved from 0.58 to 0.70 in the case of ResNet-152.
\item The explainability of the ResNet models for disease classification has been studied with LIME,  RISE, and GradCAM techniques. Studying the explainability of correct classifications helps understand if the model is making decisions based on meaningful and relevant features in the data.

\end{itemize}

The rest of the paper is organized as follows. Related work is discussed in Section \ref{RW}. Section \ref{pre} discusses the preliminaries and the proposed methodology is explained in Section \ref{PW}. Experimental results and analysis are discussed in Section \ref{ER}. Finally, conclusions are given in Section \ref{con}.

\section{Related work}
\label{RW}
Various retinal fundus image datasets have been used for training deep learning models for disease classification. One of the datasets extensively utilized is the ODIR-5K dataset which is a collection of left and right eye images of 5000 patients, publicly accessible. 

A significant challenge with the ODIR-5K dataset is the class imbalance issue, where sample counts vary significantly across different classes. This problem has not been adequately addressed by contemporary techniques, despite being a well-recognized issue in medical image classification \cite{buda2018systematic}.

Pratt et al. \cite{pratt2016convolutional} implemented a CNN-based model for the five-class classification of diabetic retinopathy, demonstrating the persistent relevance of class imbalance issues.

Islam et al. \cite{islam2019source} proposed a shallow Convolutional Neural Network (CNN)-based model, trained from scratch to classify fundus images from the ODIR-5K dataset. This model processes the left and right eye fundus images separately and assigns disease labels accordingly. Despite its simplicity, the model struggles to classify different diseases.

In contrast, Wang et al. \cite{wang2020multi} employed more sophisticated prepossessing techniques, including gray and color histogram equalization, alongside various data augmentation methods. In this paper, grey, and color histogram equalization is performed with data augmentation. 
They utilized two Efficient-Net models running in parallel on the preprocessed images, with feature concatenation occurring at the final layer for classification purposes, achieveing an AUC of 73\%.

Li et al. \cite{li2020dense} introduced a Dense Correlation Network (DCNet) based on a transfer learning ResNet model. The core component of this model is the Spatial Correlation Module (SCM), which defines dense correlations between features extracted from color fundus images at the pixel level. By fusing these correlated features, the network constructs a final feature map used to classify ocular diseases into eight categories within the ODIR-5K dataset.

Gour and Khanna \cite{gour2021multi} proposed a pre-trained, CNN architecture using two VGG-16 networks to process left and right eye fundus images in parallel. The extracted features are concatenated to form the feature map. However, despite utilizing the VGG model, they did not surpass the performance of Li et al.'s Dense Correlation Network (DCNet).

For multi-label fundus image classification, Ou et al. \cite{ou2022bfenet} suggested a two-input CNN-based attention model incorporating a multi-scale module that uses 1 × 1 and 3 × 3 dilated convolution filters. This module aims to capture multi-scale information, and a spatial attention module is employed to enhance features and establish inter dependencies between local and global information. Although computationally efficient, this model falls short of the performance achieved by Li et al.'s Dense Correlation Network (DCNet).

Sun et al. \cite{sun2022multi} proposed a hybrid graph convolution network with a LightGBM classifier, utilizing a self-attention mechanism to classify fundus images with multiple labels. Their model employed parallel CNN-based EfficientNet structures to extract features, which were subsequently concatenated and processed through a fully connected layer to determine disease labels. 
Although their model has achieved superior performance among advanced techniques, it could not detect specific illnesses within individual fundus images

Recent advancements include Xu Xia et al.'s \cite{xia2023fundus} Fundus-DeepNet, which excels in diagnosing multiple ocular diseases using deep learning but faces challenges related to image size variations and class imbalances. Similarly, Amit Bhati et al. \cite{bhati2023discriminative} introduced DKCNet, which addresses class imbalance with an attention mechanism and SE blocks, achieving superior performance with an InceptionResNet backbone. These studies underscore the ongoing need for larger datasets, enhanced deep learning architectures, and practical clinical validation.

\begin{table}[tbh]
\caption{Summary of research on Ocular Disease Classification using fundus images}
\begin{tabular}{|p{3cm}|p{4cm}|c|p{6cm}|}
\hline
\textbf{Author} & \textbf{Journal/Conf} & \textbf{Year} & \textbf{Methodology and Results} \\
\hline
Pratt et al. \cite{pratt2016convolutional}  & \multicolumn{1}{p{4cm}|}{Procedia Computer Science} & 2016 & Implemented a CNN for five-class diabetic retinopathy classification. \\
\hline
Buda et al. \cite{buda2018systematic}. & \multicolumn{1}{p{4cm}|}{Neural Networks} & 2018 & Highlighted the issue of class imbalance in medical image classification. \\
\hline
Islam et al. \cite{islam2019source} & \multicolumn{1}{p{4cm}|}{IEEE International Conference on Signal Processing (SPICSCON)} & 2019 & Proposed a shallow CNN-based model trained from scratch for classifying fundus images but struggled with disease differentiation. \\
\hline
Wang et al. \cite{wang2020multi} & \multicolumn{1}{p{4cm}|}{IEEE Access} & 2020 & Applied grey and color histogram equalization and data augmentation; used parallel EfficientNet models, achieving AUC of 73\% and F1-Score of 88\%. \\
\hline
Li et al. \cite{li2020dense} & \multicolumn{1}{p{4cm}|}{IEEE International Symposium on Biomedical Imaging (ISBI)} & 2020 & Introduced Dense Correlation Network (DCNet) with spatial correlation module (SCM) achieving 93\% AUC and 91.3\% F1-score. \\
\hline
Gour and Khanna \cite{gour2021multi} & \multicolumn{1}{p{4cm}|}{Biomedical Signal Processing and Control} & 2021 & Used pre-trained, two-input CNN with VGG-16 networks but did not outperform Li et al.'s results. \\
\hline
Ou et al. \cite{ou2022bfenet} & \multicolumn{1}{p{4cm}|}{Computer Methods and Programs in Biomedicine} & 2022 & Proposed two-input CNN-based attention model with multi-scale and spatial attention modules, yet fell short compared to Li et al.'s DCNet. \\
\hline
ASun et al. \cite{sun2022multi} & \multicolumn{1}{p{4cm}|}{Computers in Biology and Medicine} & 2022 & Proposed hybrid graph convolution network with LightGBM classifier but struggled with detecting illness in specific images. \\
\hline
Xu Xia et al. \cite{xia2023fundus}  & \multicolumn{1}{p{4cm}|}{Signal Processing: Image Communication} & 2023 & Introduced Fundus-DeepNet, excelling in diagnosing multiple ocular diseases but facing challenges like image size variations and class imbalances. \\
\hline
Bhati et al. \cite{bhati2023discriminative}  & \multicolumn{1}{p{4cm}|}{Computers in Biology and Medicine} & 2023 & Introduced DKCNet, addressing class imbalance with an attention mechanism and SE blocks, achieving superior performance with InceptionResnet backbone. \\
\hline
\end{tabular}
\label{table:research_summary}
\end{table}

\section{Preliminaries}
\label{pre}
\subsection{Normal Convolution (2D) }
\label{AA2}
\begin{figure}
     \centering
     \begin{subfigure}[b]{0.45\textwidth}
         \centering
         \includegraphics[width=\textwidth]{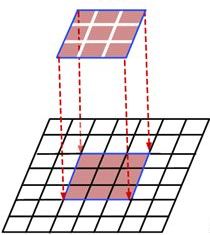}
         \caption{Normal Convolution}
         \label{fig:y equals x}
     \end{subfigure}
     \hfill
     \begin{subfigure}[b]{0.45\textwidth}
         \centering
         \includegraphics[width=\textwidth]{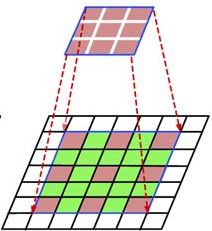}
         \caption{Dilated Convolution}
         \label{fig:three sin x}
     \end{subfigure}
     \caption{Normal convolution vs Dilated convolution in 2D}
        \label{fig:2D}
\end{figure}
Normal convolution in 2D involves sliding a filter (kernel) over the input image or feature map in both the horizontal and vertical directions. At each position, the filter is multiplied element-wise with the corresponding region of the input, and the results are summed to produce a single output value for that position. This process is performed for every position in the input, resulting in an output feature map.
Figure \ref{fig:2D}(a) shows normal convolution.
The mathematical equation for normal convolution in two-dimension is given below,

\begin{equation}
(f*g)(t,u) = \sum_{\tau= -\infty}^ {+\infty} \sum_{\varphi= -\infty}^ {+\infty} f(\tau,\varphi). g(t-\tau,u-\varphi)
\end{equation}

\subsection{Dilated Convolution (2D)}
Dilated convolution in 2D extends the receptive field of normal convolution by introducing gaps between the kernel elements in both the horizontal and vertical directions. These gaps are determined by dilation factors, which specify how far apart the kernel elements are spaced.
Figure \ref{fig:2D}(b) shows dilated convolution.
The mathematical equation for dilated convolution in two-dimension is given below,

\begin{equation}
(f*lg)(t,u) = \sum_{\tau= -\infty}^ {+\infty} \sum_{\varphi= -\infty}^ {+\infty} f(\tau,\varphi). g(t-l\tau,u-l\varphi)
\end{equation}
where $l$ is the dilation factor. If 
$l=1$; then it is normal convolution.
\subsection{Advantages of Dilated Convolution}

The proposed approach is based on dilated convolutions in the ResNet model also known as dilated ResNet. Unlike the normal ResNet which consists of normal convolution and is slow, the proposed model consists of dilated convolutions.
The advantages of dilation convolution are listed below,
\begin{itemize}
\item Increase in dilation rate of the filter will increase its receptive field.
\item Helps capture more global context from input without increasing the size of parameters along with lesser memory and computation time.
\end{itemize}
\subsection{Explainable AI(XAI)}
Explainable AI (XAI) techniques aim to find the decision-making process of AI models, particularly in complex tasks such as medical image analysis. By providing interpretable explanations, XAI methods enhance transparency and trust in AI systems, enabling users to understand the rationale behind model predictions. XAI techniques can be used to generate activation maps that highlight the areas where the model focuses on when making a classification decision. This visualization helps in understanding the regions of interest for different classes.

Various XAI techniques used in this work are explained below.
\begin{itemize}
\item  \textbf{LIME:} (Local Interpretable Model-agnostic Explanations) has been introduced by Ribeiro et al. in 2016 \cite{lime}. It approximates complex model decisions around specific instances by locally fitting interpretable models, providing intuitive explanations.
\item  \textbf{RISE:} (Randomized Input Sampling for Explanation) has been introduced by Petsiuk et al. in 2018 \cite{rise}. It generates pixel-wise explanations by systematically occluding random image patches, providing fine-grained insights into the importance of different image regions for model predictions.
 \item \textbf{Grad-CAM:} (Gradient-weighted Class Activation Mapping) has been proposed by Selvaraju et al. in 2017 \cite{gradcam}. It visualizes relevant image regions by computing gradients of class scores to convolutional feature maps, offering insights into the model's decision-making process.
\end{itemize}

\subsection{Activation Maps}
Activation maps are used to visualize and interpret the features that a neural network, particularly convolutional neural networks (CNNs), learns at various layers. They provide insight into the model's decision-making process by highlighting which regions of an input image contribute most to its predictions. This aids in debugging and improving model performance by identifying under performing layers or irrelevant features. Additionally, activation maps allow for the comparison of different model architectures, such as normal convolution versus dilated convolution, and help in understanding model sensitivity and robustness to input variations.

\section{Proposed Methodology}
\label{PW}
\begin{figure*}[htbp]
\centerline{\includegraphics[scale=.5]{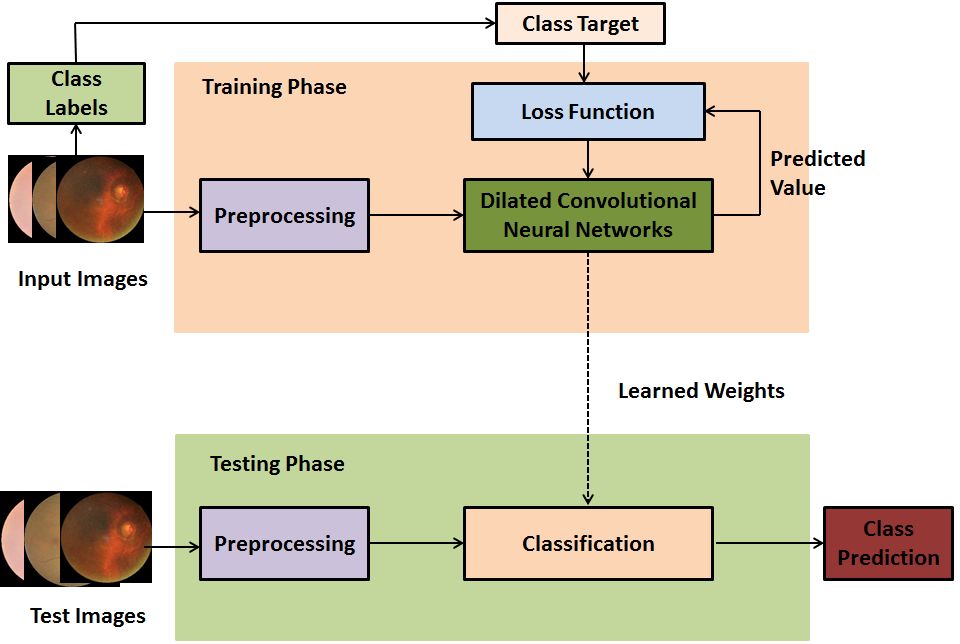}}
\caption{Flow diagram of the proposed work}
\label{fig:flow}
\end{figure*}

The flow diagram of the proposed work is shown in Figure \ref{fig:flow}.
It consists of two important phases: training and testing.
Training consists of three modules namely, i) preprocessing, ii) dilated CNN, and iii) loss function. During testing, the learned weights are used to classify the disease images.
Each of these modules is explained in detail below.
\begin{figure*}[htbp]
\centerline{\includegraphics[scale=0.35]{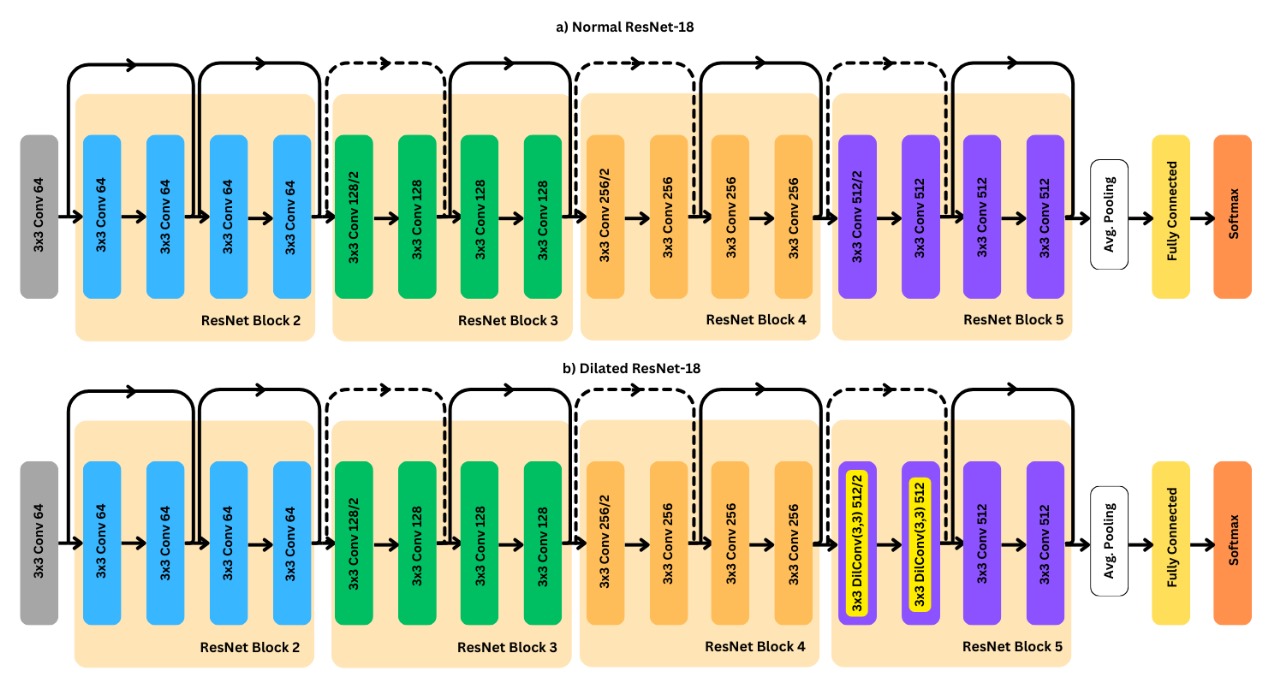}}
\caption{ResNet-18 model (a) Normal convolution (b) Dilated convolution}
\label{fig:R18}
\end{figure*}

\begin{figure*}[htbp]
\centerline{\includegraphics[scale=0.4]{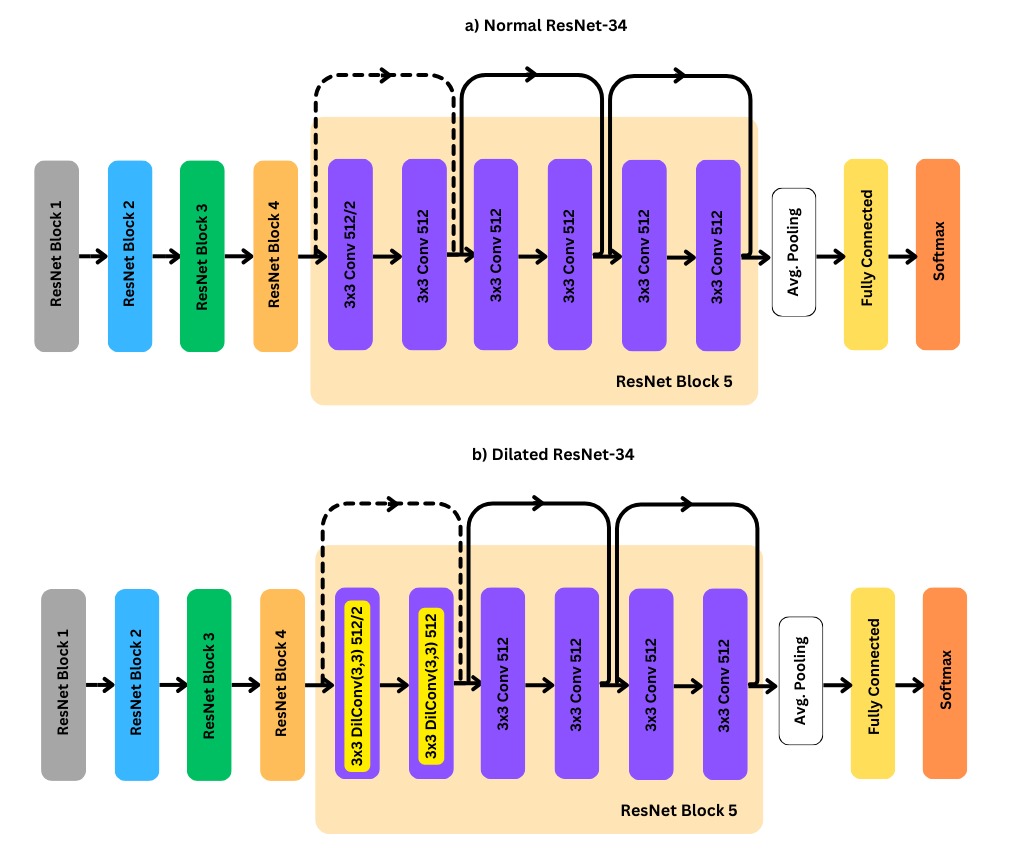}}
\caption{ResNet-34 model (a) Normal convolution (b) Dilated convolution}
\label{fig:R34}
\end{figure*}

\begin{figure*}[htbp]
\centerline{\includegraphics[scale=0.4]{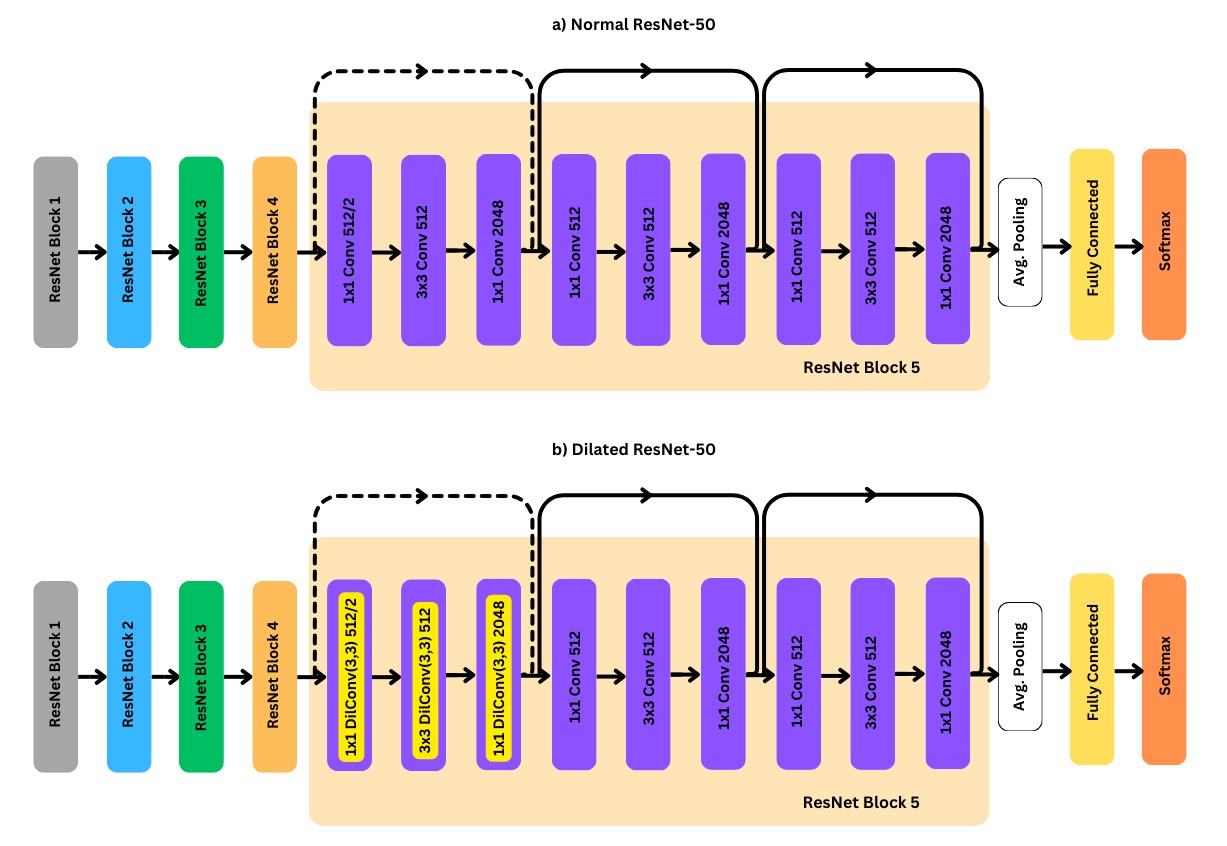}}
\caption{ResNet-50 model (a) Normal convolution (b) Dilated convolution}
\label{fig:R50}
\end{figure*}

\subsection{Preprocessing}
The retina from ODIR-5K dataset is captured using various cameras with different resolutions. Therefore, it is resized to $224 \times 224$ for preprocessing. Additionally, all the images are assigned an eight category vector based on their labels to train the models.

\subsection{Dilated CNN: Dilated ResNet Model}

The proposed model is based on dilated ResNet over normal ResNet.
Five different versions of ResNet models of varying network depth have been used in the study, namely ResNet-18, ResNet-34, ResNet-50, ResNet-101, and ResNet-152.
ResNet-18 consists of 18 layers, which include convolutional layers, pooling layers, and fully connected layers. Specifically, it has 17 convolutional layers and one fully connected layer at the end.
The first layer is a convolutional layer with 64 filters of size 7x7, followed by a max-pooling layer.
This is followed by four groups of residual blocks. Each group contains two residual blocks.
The network ends with a global average pooling layer and a fully connected layer with a softmax activation function for classification.
The difference between the normal and dilated ResNet model for 18-layer depth is shown in Figure \ref{fig:R18}. 
It can be observed that dilated convolutions with a dilation rate of (3,3) and a kernel size of (3x3) are added in the fifth block (last block) at the first two convolution layers.

The normal and dilated ResNet model for 34 layer depth is shown in Figure \ref{fig:R34}. It can be observed that dilated convolutions are added in the last block (fifth block) at the first two convolution layers. 
Figure \ref{fig:R50} shows the normal and dilated ResNet model for 50 layer depth. In these versions, dilated convolutions are added in the last block (fifth block) at the first three convolution layers.
In all the models, dilated convolutions are added in the last block (fifth block) to improve the receptive field and gather features.
The purpose of adding dilated convolutions in the higher layer blocks of ResNet is listed below,

\begin{itemize}
\item In deep neural networks, small output feature maps are obtained at the end of the network due to multiple convolution and pooling layers.
\item A large output feature map is desired for better classification by the fully connected network.
\item Remove striding in the network in order to increase the resolution of feature map reduces the receptive field which severely reduces the amount of context.
\item For this reason, dilated convolutions are used to increase the receptive field of the higher layers, compensating for the reduction in receptive field induced by removing striding.
\item This increases the receptive field without reducing spatial resolution.
\end{itemize}

So, in the last blocks of ResNet, adding dilated convolutions can help in capturing a broader context of the input image while preserving the spatial resolution, which is crucial for tasks like segmentation, classification, or any other task where dense predictions are needed.

These advantages make dilated convolutions a valuable enhancement in modern deep-learning models, especially for applications requiring detailed spatial understanding.


\subsection{Loss Function}
\begin{figure}[htbp]
\centerline{\includegraphics[scale=.3]{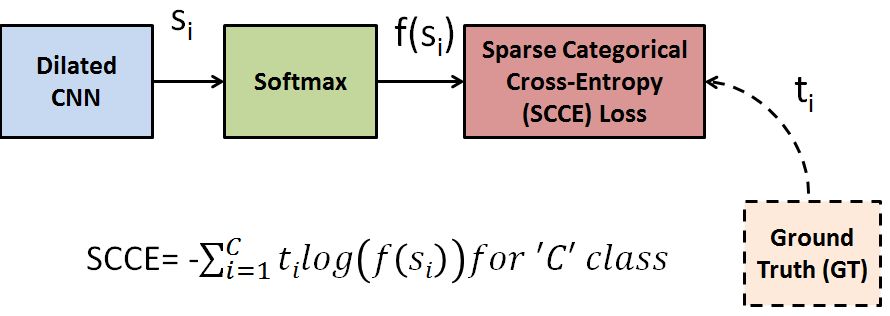}}
\caption{Sparse Categorical Cross Entropy (SCCE) loss}
\label{fig:loss}
\end{figure}
The loss function used in the proposed approach is the Sparse Categorical Cross Entropy (SCCE).
Sparse Categorical Cross Entropy (SCCE) loss function is shown in Figure  \ref{fig:loss}. The formula for the softmax function is given below:

\begin{equation}
f(s_i) = \frac{e^s_i}{\sum_{j=1}^ {C}  e^s_j}
\end{equation}

The output of the softmax is given as input to the Sparse Categorical Cross Entropy (SCCE). SCCE is the loss function that computes the difference between the actual value $f(s_i)$ with the ground truth $t_i$

\begin{equation}
SCCE = - \sum_{i=1}^ {C} t_i log (f(s_i))
\end{equation}






The advantage of using SCCE is that it is more efficient for multiple categories or classes. Unlike SCCE, the CCE (Categorical Cross Entropy) would consume a huge amount of RAM if one-hot encoded When the categories or classes are more. Also, SCCE saves memory as well as speed up the computation process.

\section{Experimental Results}
\label{ER}
\subsection{Dataset}
\par
\begin{figure*}[htbp]
\centerline{\includegraphics[scale=0.75]{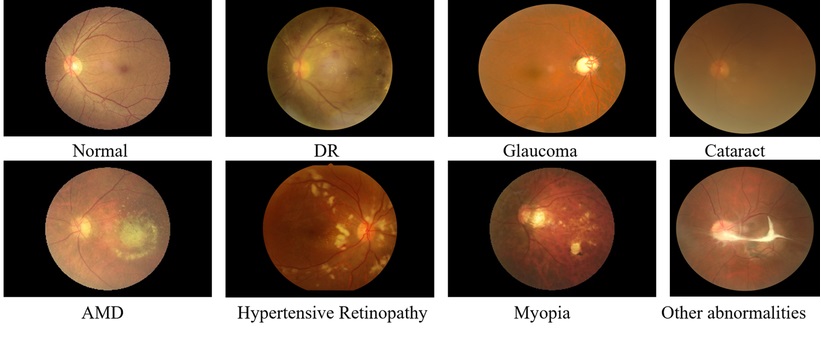}}
\caption{Sample images from ODIR dataset}
\label{fig:ODIR}
\end{figure*}

In this experiment, the ODIR-5K multi-label retinal image dataset \cite{ODIR} is used. 
Some sample images are shown in Figure \ref{fig:ODIR}. This dataset contains a set of fundus images collected from patients by Shanggong Medical Technology Co. Ltd. from different hospitals/medical centers in China. 

The original dataset contains about 7000 training and 3000 testing images. In this work though, the authors have used only the 7000 publicly available training image set for experimentation since a number of images from the test image set have been hidden as a part of the onsite challenge conducted.

\begin{figure*}[htbp]
\centerline{\includegraphics[scale=0.7]{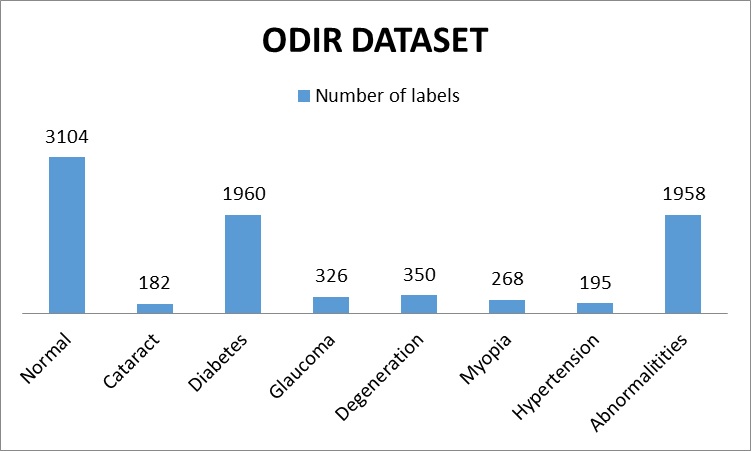}}
\caption{Class distributions of ODIR dataset}
\label{fig:distribution}
\end{figure*}

The 7,000 fundus images used in this work are taken from 3,500 patients, with both right and left eye images provided for each patient. 
Each image is assigned multiple labels, also known as multi-class labels or ground truth. Therefore, the dataset contains 7,000 fundus images with a total of 8,343 labels (as some images have multiple labels), belonging into eight different classes.
It is to be noted that, multi labeled dataset leads to a long tail distribution problem \cite{rodriguez2022multi}. 
The class wise distribution of the dataset is shown in Figure \ref{fig:distribution}. This dataset has eight classes labeled as normal, cataract, diabetic retinopathy (DR), glaucoma, age related macular degeneration (AMD), myopia, hypertension, and other abnormalities. 
There is a slight imbalance in the dataset as well. The sample images are more for the classes such as normal, diabetic retinopathy, and other abnormalities. The sample images for hypertension and cataract are less in number.

\subsection{Metrics Used }\label{AA10}
Metrics such as accuracy, precision, recall, and F1 score are used to evaluate the proposed methodology.

\begin{itemize}

\item
Accuracy is the ratio of correctly predicted instances to the total instances. It is calculated as:

\begin{equation}
Accuracy = \frac{TP + TN}{TP + TN + FP + FN}
\end{equation}

where:
\begin{itemize}
    \item $TP$ (True Positive) is the number of positive instances correctly classified.
    \item $TN$ (True Negative) is the number of negative instances correctly classified.
    \item $FP$ (False Positive) is the number of negative instances incorrectly classified as positive.
    \item $FN$ (False Negative) is the number of positive instances incorrectly classified as negative.
\end{itemize}

\item
Precision, also known as positive predictive value, is the ratio of correctly predicted positive observations to the total predicted positives. It is calculated as:

\begin{equation}
Precision = \frac{TP}{TP + FP}
\end{equation}

\item
Recall, also known as sensitivity or true positive rate, is the ratio of correctly predicted positive observations to all observations in the actual class. It is calculated as:

\begin{equation}
Recall = \frac{TP}{TP + FN}
\end{equation}

\item
The F1 score is the harmonic mean of precision and recall. It is calculated as:

\begin{equation}
F1 Score = 2 * \frac{Precision * Recall}{Precision + Recall} \label{eq4}
\end{equation}

\end{itemize}


\subsection{Results}
Table \ref{Resnet-18} shows the results of the ResNet-18 model without and with dilation trained and tested on the ODIR dataset. Two models were used in this experiment - the base model (without dilation) and the other one with dilation.
Results containing precision, recall, and F1 score are given in Table \ref{Resnet-18}.
The overall F1 score without dilation is 0.69 and with dilation is 0.71.
It can be noticed that the F1 score has improved for the classes, namely glaucoma, degeneration, myopia, and hypertension. The overall accuracy obtained is 0.78 for both without and with dilation as given in Table \ref{overall}.

Similarly, Table \ref{Resnet-34} shows the results obtained on the same dataset using the ResNet-34 model with and without dilation. The overall F1 score without dilation is 0.68 and with dilation is 0.70. In terms of accuracy as well, there is a marked improvement from 0.77 to 0.79 as given in Table \ref{overall}. 
It can be noticed that the F1 score has improved for the classes, namely normal, diabetes, and abnormalities.

Table \ref{Resnet-50} shows the results obtained using the ResNet-50 model with and without dilation. The overall F1 score with and without dilation is 0.69. In terms of accuracy, there is a marked improvement from 0.76 to 0.77 as given in Table \ref{overall}. 
It can be noticed that the F1 score has improved for the classes, namely glaucoma, degeneration, and abnormalities.

Table \ref{Resnet-101} shows the results obtained using the ResNet-101 model with and without dilation. 
The overall F1 score without dilation is 0.54 and with dilation is 0.67. In terms of accuracy as well, there is a marked improvement from 0.62 to 0.76 as given in Table \ref{overall}. 
It can be observed that the F1 score has improved for all eight classes with dilation as compared to those without dilation for deeper (as in case of Resnet-101) models.

Similarly, Table \ref{Resnet-152} shows the results obtained using the ResNet-152 model for with and without dilation. 
The overall F1 score without dilation is 0.58 and with dilation is 0.70. In terms of accuracy as well, there is a marked improvement from 0.70 to 0.79 as given in Table \ref{overall}.
It can be observed that the F1 score has improved for all eight classes with dilation as compared to those without dilation for very deep (as in case of Resnet-152) models.

The average accuracy and F1 score obtained for all five models are given in Table \ref{overall}.
Overall, the results presented in Tables \ref{Resnet-18}, \ref{Resnet-34}, \ref{Resnet-50}, \ref{Resnet-101}, and \ref{Resnet-152} show a clear trend where the use of dilation generally enhances the performance of ResNet models on the ODIR dataset. Across different models, the introduction of dilation improved the overall F1 score and accuracy, with notable improvements in specific classes such as glaucoma, degeneration, myopia, hypertension, normal, diabetes, and abnormalities. For instance, ResNet-18 and ResNet-34 showed modest increases in F1 scores and slight improvements in accuracy, while deeper models like ResNet-101 and ResNet-152 demonstrated significant enhancements in both F1 scores (increasing by 0.13 and 0.12, respectively) and accuracy (increasing by 0.14 and 0.09, respectively). This suggests that deeper architectures benefit more from dilation, due to more reduction in feature map sizes at the end of the network. Overall, the use of dilation consistently resulted in better model performance, highlighting its effectiveness in improving precision, recall, and accuracy in medical image classification tasks.

\begin{table}[tbh]
\resizebox{\columnwidth}{!}{
\begin{tabular}{|l|cccc|cccc|}
\hline
\multicolumn{1}{|c|}{\multirow{2}{*}{\textbf{Disease}}} & \multicolumn{4}{c|}{\textbf{ResNet-18 without Dilation}} & \multicolumn{4}{c|}{\textbf{ResNet-18 with Dilation}} \\
\multicolumn{1}{|c|}{} & \multicolumn{1}{l|}{\textbf{Precision}} & \multicolumn{1}{l|}{\textbf{Recall}} & \multicolumn{1}{l|}{\textbf{F1-Score}} & \multicolumn{1}{l|}{\textbf{Support}} & \multicolumn{1}{l|}{\textbf{Precision}} & \multicolumn{1}{l|}{\textbf{Recall}} & \multicolumn{1}{l|}{\textbf{F1-Score}} & \multicolumn{1}{l|}{\textbf{Support}} \\ \hline
cataract & \multicolumn{1}{c|}{0.75} & \multicolumn{1}{c|}{0.95} & \multicolumn{1}{c|}{0.84} & 183 & \multicolumn{1}{c|}{0.75} & \multicolumn{1}{c|}{0.97} & \multicolumn{1}{c|}{0.84} & 183 \\ \hline
normal & \multicolumn{1}{c|}{0.80} & \multicolumn{1}{c|}{0.89} & \multicolumn{1}{c|}{0.84} & 1276 & \multicolumn{1}{c|}{0.88} & \multicolumn{1}{c|}{0.80} & \multicolumn{1}{c|}{0.84} & 1276 \\ \hline
glaucoma & \multicolumn{1}{c|}{0.67} & \multicolumn{1}{c|}{0.68} & \multicolumn{1}{c|}{0.67} & 182 & \multicolumn{1}{c|}{0.66} & \multicolumn{1}{c|}{0.75} & \multicolumn{1}{c|}{\textbf{0.70}} & 182 \\ \hline
diabetes & \multicolumn{1}{c|}{0.81} & \multicolumn{1}{c|}{0.74} & \multicolumn{1}{c|}{0.77} & 1047 & \multicolumn{1}{c|}{0.73} & \multicolumn{1}{c|}{0.81} & \multicolumn{1}{c|}{0.77} & 1047 \\ \hline
degeneration & \multicolumn{1}{c|}{0.82} & \multicolumn{1}{c|}{0.65} & \multicolumn{1}{c|}{0.72} & 161 & \multicolumn{1}{c|}{0.81} & \multicolumn{1}{c|}{0.68} & \multicolumn{1}{c|}{\textbf{0.74}} & 161 \\ \hline
myopia & \multicolumn{1}{c|}{0.83} & \multicolumn{1}{c|}{0.97} & \multicolumn{1}{c|}{0.90} & 132 & \multicolumn{1}{c|}{0.93} & \multicolumn{1}{c|}{0.97} & \multicolumn{1}{c|}{\textbf{0.95}} & 132 \\ \hline
hypertension & \multicolumn{1}{c|}{0.43} & \multicolumn{1}{c|}{0.61} & \multicolumn{1}{c|}{0.50} & 90 & \multicolumn{1}{c|}{0.47} & \multicolumn{1}{c|}{0.67} & \multicolumn{1}{c|}{\textbf{0.55}} & 90 \\ \hline
abnormalities & \multicolumn{1}{c|}{0.94} & \multicolumn{1}{c|}{0.20} & \multicolumn{1}{c|}{0.33} & 154 & \multicolumn{1}{c|}{0.72} & \multicolumn{1}{c|}{0.20} & \multicolumn{1}{c|}{0.31} & 154 \\ \hline
Avg. F1  & \multicolumn{1}{l|}{} & \multicolumn{1}{l|}{} & \multicolumn{1}{c|}{0.69} &  & \multicolumn{1}{l|}{} & \multicolumn{1}{l|}{} & \multicolumn{1}{c|}{\textbf{0.71}} &  \\ \hline
\end{tabular}
}
\caption{Results of ResNet-18 with and without dilation}
\label{Resnet-18}
\end{table}

\begin{table}[tbh]
\resizebox{\columnwidth}{!}{%
\begin{tabular}{|l|cccc|cccc|}
\hline
\multicolumn{1}{|c|}{\multirow{2}{*}{\textbf{Disease}}} & \multicolumn{4}{c|}{\textbf{ResNet-34 without Dilation}} & \multicolumn{4}{c|}{\textbf{ResNet-34 with Dilation}} \\ 
\multicolumn{1}{|c|}{} & \multicolumn{1}{l|}{\textbf{Precision}} & \multicolumn{1}{l|}{\textbf{Recall}} & \multicolumn{1}{l|}{\textbf{F1-Score}} & \multicolumn{1}{l|}{\textbf{Support}} & \multicolumn{1}{l|}{\textbf{Precision}} & \multicolumn{1}{l|}{\textbf{Recall}} & \multicolumn{1}{l|}{\textbf{F1-Score}} & \multicolumn{1}{l|}{\textbf{Support}} \\ \hline
cataract & \multicolumn{1}{c|}{0.87} & \multicolumn{1}{c|}{0.88} & \multicolumn{1}{c|}{0.88} & 183 & \multicolumn{1}{c|}{0.85} & \multicolumn{1}{c|}{0.92} & \multicolumn{1}{c|}{0.88} & 183 \\ \hline
normal & \multicolumn{1}{c|}{0.87} & \multicolumn{1}{c|}{0.80} & \multicolumn{1}{c|}{0.83} & 1276 & \multicolumn{1}{c|}{0.82} & \multicolumn{1}{c|}{0.88} & \multicolumn{1}{c|}{\textbf{0.85}} & 1276 \\ \hline
glaucoma & \multicolumn{1}{c|}{0.58} & \multicolumn{1}{c|}{0.81} & \multicolumn{1}{c|}{0.68} & 182 & \multicolumn{1}{c|}{0.65} & \multicolumn{1}{c|}{0.72} & \multicolumn{1}{c|}{0.68} & 182 \\ \hline
diabetes & \multicolumn{1}{c|}{0.70} & \multicolumn{1}{c|}{0.85} & \multicolumn{1}{c|}{0.77} & 1047 & \multicolumn{1}{c|}{0.76} & \multicolumn{1}{c|}{0.79} & \multicolumn{1}{c|}{\textbf{0.78}} & 1047 \\ \hline
degeneration & \multicolumn{1}{c|}{0.90} & \multicolumn{1}{c|}{0.47} & \multicolumn{1}{c|}{0.62} & 161 & \multicolumn{1}{c|}{0.85} & \multicolumn{1}{c|}{0.61} & \multicolumn{1}{c|}{\textbf{0.71}} & 161 \\ \hline
myopia & \multicolumn{1}{c|}{0.93} & \multicolumn{1}{c|}{0.95} & \multicolumn{1}{c|}{0.94} & 132 & \multicolumn{1}{c|}{0.90} & \multicolumn{1}{c|}{0.96} & \multicolumn{1}{c|}{0.93} & 132 \\ \hline
hypertension & \multicolumn{1}{c|}{0.47} & \multicolumn{1}{c|}{0.50} & \multicolumn{1}{c|}{0.49} & 90 & \multicolumn{1}{c|}{0.45} & \multicolumn{1}{c|}{0.48} & \multicolumn{1}{c|}{0.46} & 90 \\ \hline
abnormalities & \multicolumn{1}{c|}{0.93} & \multicolumn{1}{c|}{0.17} & \multicolumn{1}{c|}{0.29} & 154 & \multicolumn{1}{c|}{0.89} & \multicolumn{1}{c|}{0.21} & \multicolumn{1}{c|}{\textbf{0.35}} & 154 \\ \hline
Avg. F1  & \multicolumn{1}{l|}{} & \multicolumn{1}{l|}{} & \multicolumn{1}{c|}{0.68} & & \multicolumn{1}{l|}{} & \multicolumn{1}{l|}{} & \multicolumn{1}{c|}{\textbf{0.70}} &  \\ \hline
\end{tabular}%
}
\caption{Results of ResNet-34 with and without dilation}
\label{Resnet-34}
\end{table}

\begin{table}[]
\resizebox{\columnwidth}{!}{%
\begin{tabular}{|l|cccc|cccc|}
\hline
\multicolumn{1}{|c|}{\multirow{2}{*}{\textbf{Disease}}} & \multicolumn{4}{c|}{\textbf{ResNet-50 without Dilation}} & \multicolumn{4}{c|}{\textbf{ResNet-50 with Dilation}} \\ 
\multicolumn{1}{|c|}{} & \multicolumn{1}{l|}{\textbf{Precision}} & \multicolumn{1}{l|}{\textbf{Recall}} & \multicolumn{1}{l|}{\textbf{F1-Score}} & \multicolumn{1}{l|}{\textbf{Support}} & \multicolumn{1}{l|}{\textbf{Precision}} & \multicolumn{1}{l|}{\textbf{Recall}} & \multicolumn{1}{l|}{\textbf{F1-Score}} & \multicolumn{1}{l|}{\textbf{Support}} \\ \hline
cataract & \multicolumn{1}{c|}{0.82} & \multicolumn{1}{c|}{0.91} & \multicolumn{1}{c|}{0.86} & 183 & \multicolumn{1}{c|}{0.76} & \multicolumn{1}{c|}{0.92} & \multicolumn{1}{c|}{0.83} & 183 \\ \hline
normal & \multicolumn{1}{c|}{0.80} & \multicolumn{1}{c|}{0.89} & \multicolumn{1}{c|}{0.85} & 1276 & \multicolumn{1}{c|}{0.80} & \multicolumn{1}{c|}{0.85} & \multicolumn{1}{c|}{0.83} & 1276 \\ \hline
glaucoma & \multicolumn{1}{c|}{0.60} & \multicolumn{1}{c|}{0.74} & \multicolumn{1}{c|}{0.66} & 182 & \multicolumn{1}{c|}{0.74} & \multicolumn{1}{c|}{0.64} & \multicolumn{1}{c|}{\textbf{0.69}} & 182 \\ \hline
diabetes & \multicolumn{1}{c|}{0.87} & \multicolumn{1}{c|}{0.70} & \multicolumn{1}{c|}{0.77} & 1047 & \multicolumn{1}{c|}{0.75} & \multicolumn{1}{c|}{0.78} & \multicolumn{1}{c|}{0.76} & 1047 \\ \hline
degeneration & \multicolumn{1}{c|}{0.56} & \multicolumn{1}{c|}{0.78} & \multicolumn{1}{c|}{0.65} & 161 & \multicolumn{1}{c|}{0.85} & \multicolumn{1}{c|}{0.61} & \multicolumn{1}{c|}{\textbf{0.71}} & 161 \\ \hline
myopia & \multicolumn{1}{c|}{0.95} & \multicolumn{1}{c|}{0.90} & \multicolumn{1}{c|}{0.93} & 132 & \multicolumn{1}{c|}{0.90} & \multicolumn{1}{c|}{0.95} & \multicolumn{1}{c|}{0.92} & 132 \\ \hline
hypertension & \multicolumn{1}{c|}{0.45} & \multicolumn{1}{c|}{0.69} & \multicolumn{1}{c|}{0.54} & 90 & \multicolumn{1}{c|}{0.45} & \multicolumn{1}{c|}{0.50} & \multicolumn{1}{c|}{0.47} & 90 \\ \hline
abnormalities & \multicolumn{1}{c|}{0.63} & \multicolumn{1}{c|}{0.22} & \multicolumn{1}{c|}{0.33} & 154 & \multicolumn{1}{c|}{0.87} & \multicolumn{1}{c|}{0.21} & \multicolumn{1}{c|}{\textbf{0.34}} & 154 \\ \hline
Avg. F1  & \multicolumn{1}{l|}{} & \multicolumn{1}{l|}{} & \multicolumn{1}{c|}{0.69} & & \multicolumn{1}{l|}{} & \multicolumn{1}{l|}{} & \multicolumn{1}{c|}{\textbf{0.69}} &   \\ \hline
\end{tabular}
}
\caption{Results of ResNet-50 with and without dilation}
\label{Resnet-50}
\centering
\end{table}
\begin{table}[tbh]
\resizebox{\columnwidth}{!}{%
\begin{tabular}{|l|cccc|cccc|}
\hline
\multicolumn{1}{|c|}{\multirow{2}{*}{\textbf{Disease}}} & \multicolumn{4}{c|}{\textbf{ResNet-101 without Dilation}} & \multicolumn{4}{c|}{\textbf{ResNet-101 with Dilation}} \\  
\multicolumn{1}{|c|}{} & \multicolumn{1}{l|}{\textbf{Precision}} & \multicolumn{1}{l|}{\textbf{Recall}} & \multicolumn{1}{l|}{\textbf{F1-Score}} & \multicolumn{1}{l|}{\textbf{Support}} & \multicolumn{1}{l|}{\textbf{Precision}} & \multicolumn{1}{l|}{\textbf{Recall}} & \multicolumn{1}{l|}{\textbf{F1-Score}} & \multicolumn{1}{l|}{\textbf{Support}} \\ \hline
cataract & \multicolumn{1}{c|}{0.49} & \multicolumn{1}{c|}{0.95} & \multicolumn{1}{c|}{0.65} & 183 & \multicolumn{1}{c|}{0.81} & \multicolumn{1}{c|}{0.91} & \multicolumn{1}{c|}{\textbf{0.85}} & 183 \\ \hline
normal & \multicolumn{1}{c|}{0.68} & \multicolumn{1}{c|}{0.79} & \multicolumn{1}{c|}{0.73} & 1276 & \multicolumn{1}{c|}{0.85} & \multicolumn{1}{c|}{0.82} & \multicolumn{1}{c|}{\textbf{0.83}} & 1276 \\ \hline
glaucoma & \multicolumn{1}{c|}{0.57} & \multicolumn{1}{c|}{0.60} & \multicolumn{1}{c|}{0.58} & 182 & \multicolumn{1}{c|}{0.52} & \multicolumn{1}{c|}{0.73} & \multicolumn{1}{c|}{\textbf{0.61}} & 182 \\ \hline
diabetes & \multicolumn{1}{c|}{0.72} & \multicolumn{1}{c|}{0.43} & \multicolumn{1}{c|}{0.54} & 1047 & \multicolumn{1}{c|}{0.80} & \multicolumn{1}{c|}{0.74} & \multicolumn{1}{c|}{\textbf{0.77}} & 1047 \\ \hline
degeneration & \multicolumn{1}{c|}{0.54} & \multicolumn{1}{c|}{0.54} & \multicolumn{1}{c|}{0.54} & 161 & \multicolumn{1}{c|}{0.50} & \multicolumn{1}{c|}{0.75} & \multicolumn{1}{c|}{\textbf{0.60}} & 161 \\ \hline
myopia & \multicolumn{1}{c|}{0.96} & \multicolumn{1}{c|}{0.61} & \multicolumn{1}{c|}{0.74} & 132 & \multicolumn{1}{c|}{0.87} & \multicolumn{1}{c|}{0.95} & \multicolumn{1}{c|}{\textbf{0.91}} & 132 \\ \hline
hypertension & \multicolumn{1}{c|}{0.27} & \multicolumn{1}{c|}{0.52} & \multicolumn{1}{c|}{0.36} & 90 & \multicolumn{1}{c|}{0.41} & \multicolumn{1}{c|}{0.71} & \multicolumn{1}{c|}{\textbf{0.52}} & 90 \\ \hline
abnormalities & \multicolumn{1}{c|}{0.20} & \multicolumn{1}{c|}{0.21} & \multicolumn{1}{c|}{0.21} & 154 & \multicolumn{1}{c|}{1.00} & \multicolumn{1}{c|}{0.18} & \multicolumn{1}{c|}{\textbf{0.30}} & 154 \\ \hline
Avg. F1  & \multicolumn{1}{l|}{} & \multicolumn{1}{l|}{} & \multicolumn{1}{c|}{0.54} &  & \multicolumn{1}{l|}{} & \multicolumn{1}{l|}{} & \multicolumn{1}{c|}{\textbf{0.67}} &  \\ \hline
\end{tabular}%
}
\caption{Results of ResNet-101 with and without dilation}
\label{Resnet-101}
\end{table}

\begin{table}[tbh]
\resizebox{\columnwidth}{!}{%
\begin{tabular}{|l|cccc|cccc|}
\hline
\multicolumn{1}{|c|}{\multirow{2}{*}{\textbf{Disease}}} & \multicolumn{4}{c|}{\textbf{ResNet-152 without Dilation}} & \multicolumn{4}{c|}{\textbf{ResNet-152 with Dilation}} \\
\multicolumn{1}{|c|}{} & \multicolumn{1}{l|}{\textbf{Precision}} & \multicolumn{1}{l|}{\textbf{Recall}} & \multicolumn{1}{l|}{\textbf{F1-Score}} & \multicolumn{1}{l|}{\textbf{Support}} & \multicolumn{1}{l|}{\textbf{Precision}} & \multicolumn{1}{l|}{\textbf{Recall}} & \multicolumn{1}{l|}{\textbf{F1-Score}} & \multicolumn{1}{l|}{\textbf{Support}} \\ \hline
cataract & \multicolumn{1}{c|}{0.73} & \multicolumn{1}{c|}{0.90} & \multicolumn{1}{c|}{0.80} & 183 & \multicolumn{1}{c|}{0.84} & \multicolumn{1}{c|}{0.89} & \multicolumn{1}{c|}{\textbf{0.86}} & 183 \\ \hline
normal & \multicolumn{1}{c|}{0.86} & \multicolumn{1}{c|}{0.67} & \multicolumn{1}{c|}{0.75} & 1276 & \multicolumn{1}{c|}{0.81} & \multicolumn{1}{c|}{0.89} & \multicolumn{1}{c|}{\textbf{0.85}} & 1276 \\ \hline
glaucoma & \multicolumn{1}{c|}{0.49} & \multicolumn{1}{c|}{0.69} & \multicolumn{1}{c|}{0.57} & 182 & \multicolumn{1}{c|}{0.68} & \multicolumn{1}{c|}{0.71} & \multicolumn{1}{c|}{\textbf{0.70}} & 182 \\ \hline
diabetes & \multicolumn{1}{c|}{0.61} & \multicolumn{1}{c|}{0.87} & \multicolumn{1}{c|}{0.72} & 1047 & \multicolumn{1}{c|}{0.78} & \multicolumn{1}{c|}{0.78} & \multicolumn{1}{c|}{\textbf{0.78}} & 1047 \\ \hline
degeneration & \multicolumn{1}{c|}{0.73} & \multicolumn{1}{c|}{0.43} & \multicolumn{1}{c|}{0.54} & 161 & \multicolumn{1}{c|}{0.77} & \multicolumn{1}{c|}{0.67} & \multicolumn{1}{c|}{\textbf{0.72}} & 161 \\ \hline
myopia & \multicolumn{1}{c|}{0.97} & \multicolumn{1}{c|}{0.68} & \multicolumn{1}{c|}{0.80} & 132 & \multicolumn{1}{c|}{0.92} & \multicolumn{1}{c|}{0.90} & \multicolumn{1}{c|}{\textbf{0.91}} & 132 \\ \hline
hypertension & \multicolumn{1}{c|}{0.60} & \multicolumn{1}{c|}{0.17} & \multicolumn{1}{c|}{0.26} & 90 & \multicolumn{1}{c|}{0.52} & \multicolumn{1}{c|}{0.44} & \multicolumn{1}{c|}{\textbf{0.48}} & 90 \\ \hline
abnormalities & \multicolumn{1}{c|}{0.53} & \multicolumn{1}{c|}{0.17} & \multicolumn{1}{c|}{0.26} & 154 & \multicolumn{1}{c|}{0.95} & \multicolumn{1}{c|}{0.23} & \multicolumn{1}{c|}{\textbf{0.37}} & 154 \\ \hline
Avg. F1  & \multicolumn{1}{l|}{} & \multicolumn{1}{l|}{} & \multicolumn{1}{c|}{0.58}&  & \multicolumn{1}{l|}{} & 
\multicolumn{1}{l|}{} & \multicolumn{1}{c|}{\textbf{0.70}} &  \\ \hline
\end{tabular}%
}
\caption{Results of ResNet-152 with and without dilation}
\label{Resnet-152}
\end{table}

\begin{table}[tbh]
\resizebox{\columnwidth}{!}{%
\begin{tabular}{|c|c|c|c|c|}
\hline
\multicolumn{1}{|l|}{ResNet types} & \begin{tabular}[c]{@{}c@{}}Accuracy \\ without Dilation\end{tabular} & \begin{tabular}[c]{@{}c@{}}Accuracy \\ with Dilation\end{tabular} & \begin{tabular}[c]{@{}c@{}}F1 Score\\ (without Dilation )\end{tabular} & \begin{tabular}[c]{@{}c@{}}F1 Score\\ (with Dilation)\end{tabular} \\ \hline
ResNet-18 & 0.78 & \textbf{0.78} & 0.69 & \textbf{0.71} \\ \hline
ResNet-34 & 0.77 & \textbf{0.79} & 0.68 & \textbf{0.70} \\ \hline
ResNet-50 & 0.76 & \textbf{0.77} & 0.69 & \textbf{0.69} \\ \hline
ResNet-101 & 0.62 & \textbf{0.76} & 0.54 & \textbf{0.67} \\ \hline
ResNet-152 & 0.70 & \textbf{0.79} & 0.58 & \textbf{0.70} \\ \hline
\end{tabular}
}
\caption{Comparison of model accuracies and average F1 scores of ResNet models with and without dilation}
\label{overall}
\end{table}

\subsection{Explainable AI Techniques} 
In the context of the ResNet models trained on the ODIR dataset for classifying eye diseases, XAI can:
\begin{itemize}
\item Visualize Attention: Show which parts of the retinal images the model focuses on when making a classification (e.g., glaucoma or myopia).
\item Explain Misclassifications: Provide insights into why certain images are misclassified, helping improve model accuracy.
\item Validate Medical Relevance: Ensure that the features the model relies on align with medical knowledge, enhancing clinician trust in the model.
\end{itemize}


\begin{figure}[H]
     \centering
     \begin{subfigure}[b]{0.45\textwidth}
         \centering
         \includegraphics[width=\textwidth]{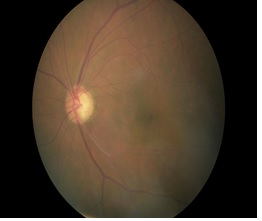}
         \caption{Input image (Glaucoma)}
     \end{subfigure}
     \hfill
     \begin{subfigure}[b]{0.45\textwidth}
         \centering
         \includegraphics[width=\textwidth]{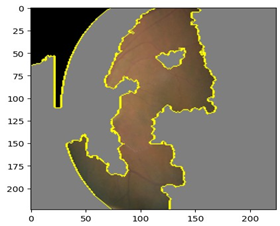}
         \caption{Explanation for other classes}
     \end{subfigure}
     \vfill
     \begin{subfigure}[b]{0.45\textwidth}
         \centering
         \includegraphics[width=\textwidth]{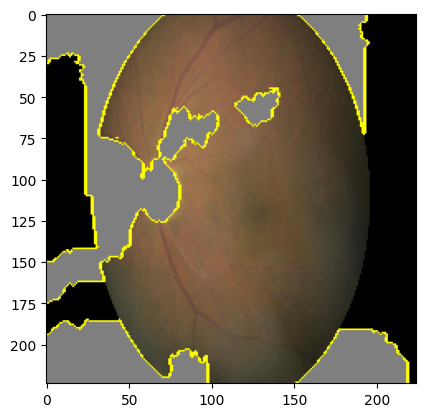}
         \caption{Without dilation: Explanation for glaucoma}
     \end{subfigure}
     \hfill
     \begin{subfigure}[b]{0.45\textwidth}
         \centering
         \includegraphics[width=\textwidth]{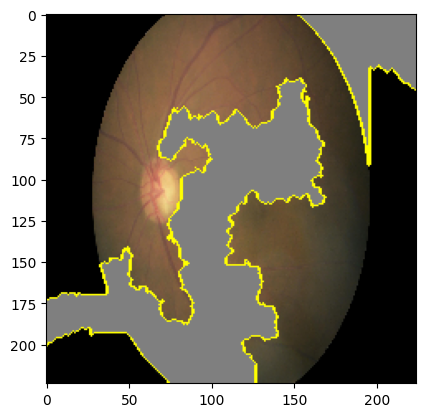}
         \caption{With dilation: Explanation for glaucoma}
     \end{subfigure}
     \caption{LIME}
     \label{fig:LIME}
\end{figure}

Figure \ref{fig:LIME} illustrates the explanation by LIME 
for the input image whose label is glaucoma shown in  Figure \ref{fig:LIME}(a). 
LIME offers local explanations by perturbing input data for individual predictions.
Figure \ref{fig:LIME}(b) highlights the explanation for other class while Figure \ref{fig:LIME}(c) and Figure \ref{fig:LIME}(d) highlight the explanation for without and with dilation respectively for the disease glaucoma.
These visualizations highlight the areas where the model focuses to make its decision, offering valuable insights for each class. It can be noted that the explanation by LIME with dilation is focused on the optic cup and disc area which is crucial for glaucoma detection while it is missing by the explanation of LIME without dilation. 
Hence, the ResNet-152 model with dilation has achieved higher accuracy in predicting the correct classes compared to the model without dilation.

\begin{figure}[H]
     \centering
     \begin{subfigure}[b]{0.45\textwidth}
         \centering
         \includegraphics[width=\textwidth]{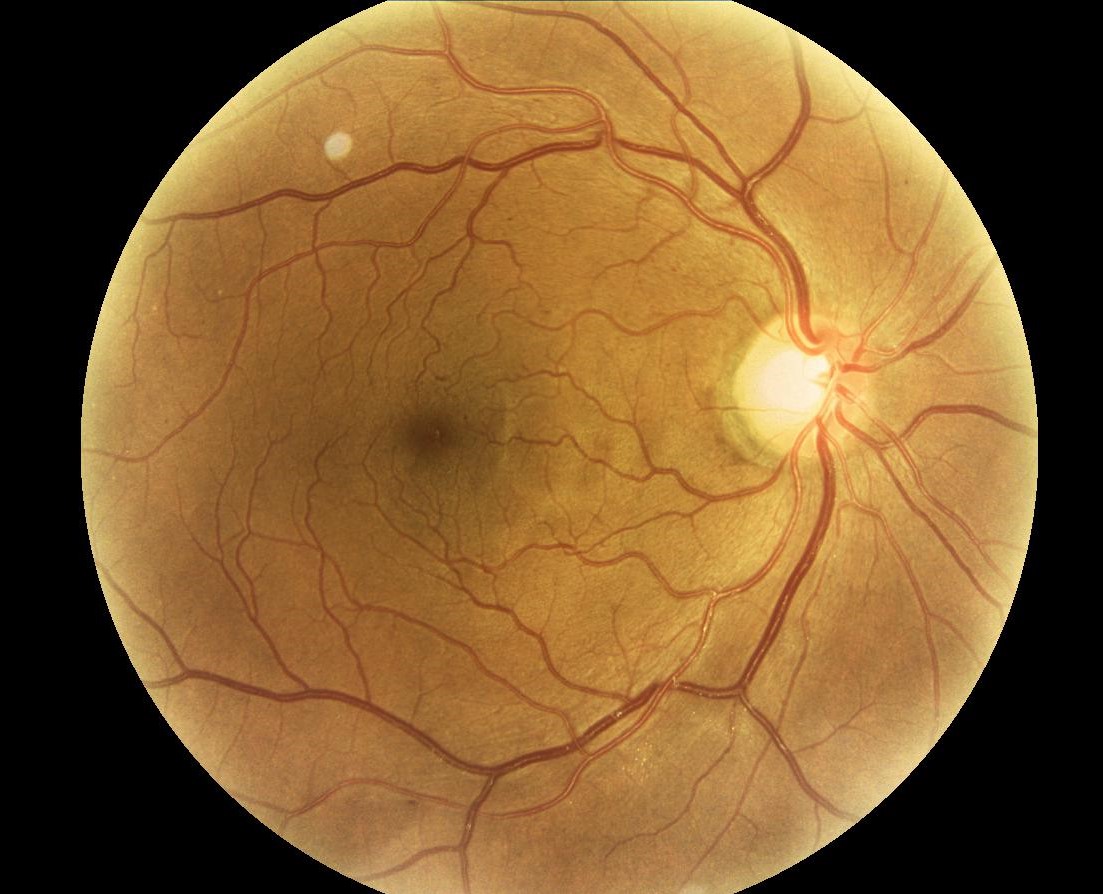}
         \caption{Input image (Diabetes)}
     \end{subfigure}
     \hfill
          \begin{subfigure}[b]{0.45\textwidth}
         \centering
         \includegraphics[width=\textwidth]{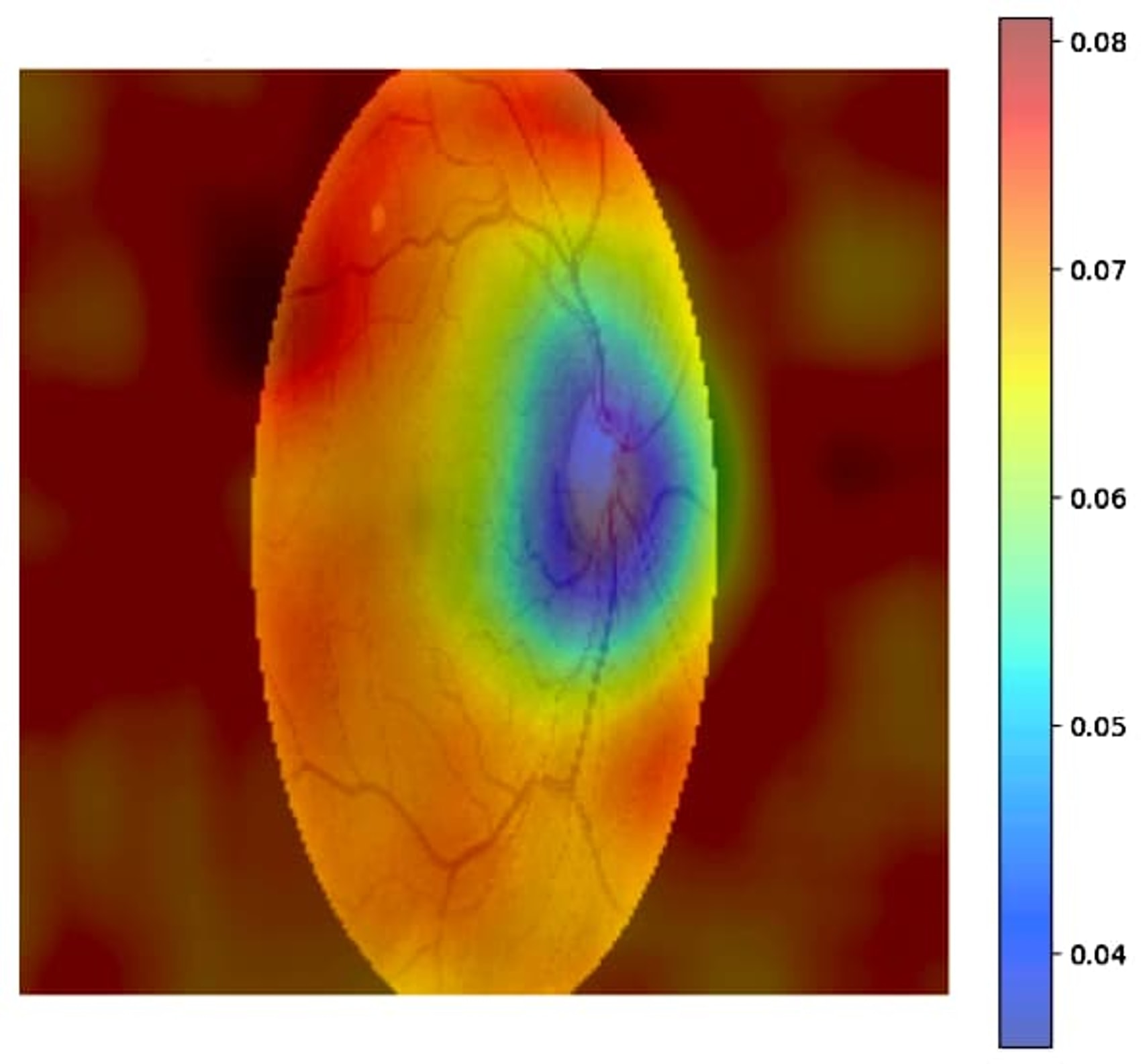}
         \caption{Explanation for other classes}
     \end{subfigure}

     \vfill
     \begin{subfigure}[b]{0.45\textwidth}
         \centering
         \includegraphics[width=\textwidth]{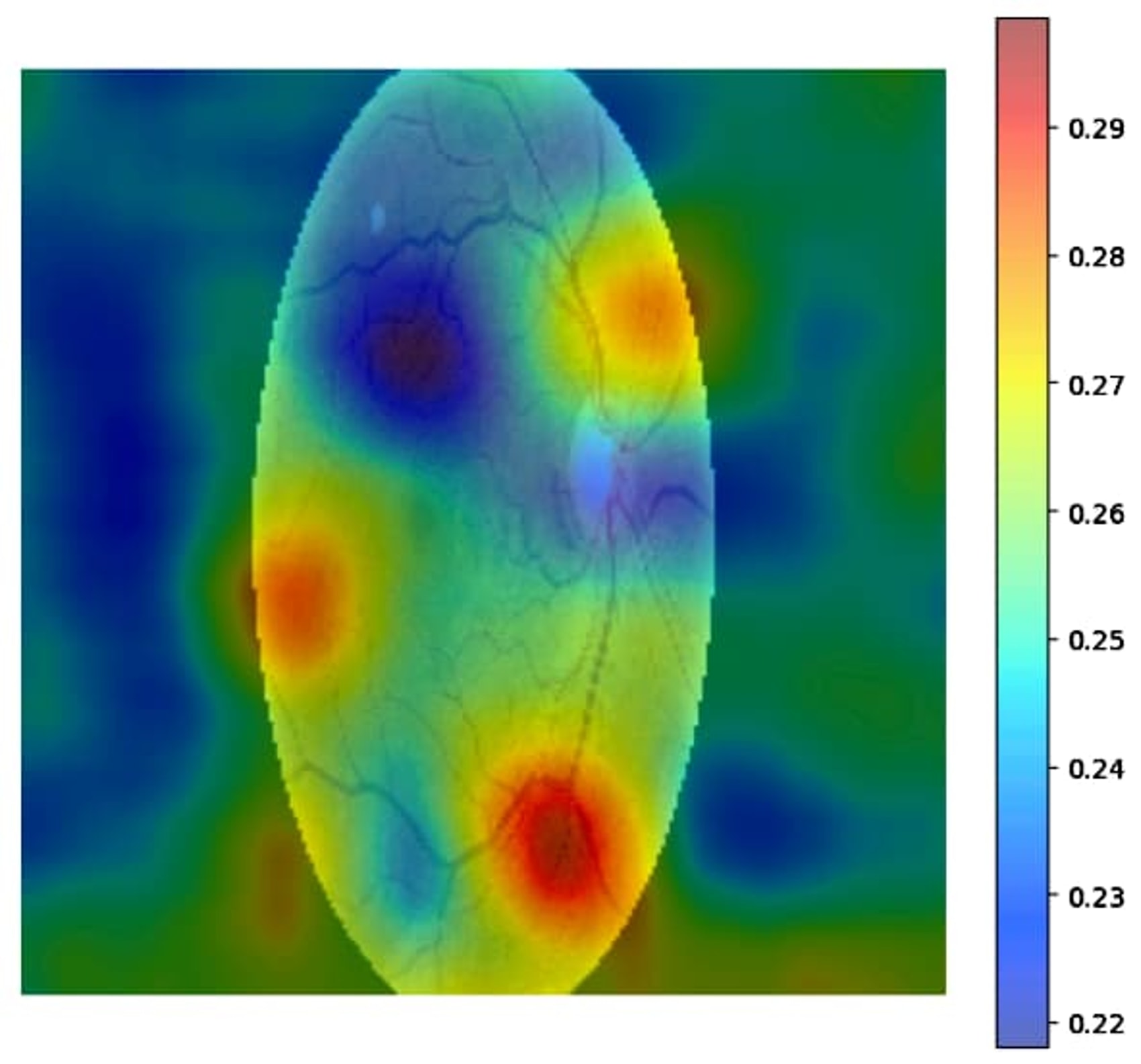}
         \caption{Without dilation: Explanation for diabetes}
     \end{subfigure}
     \hfill
     \begin{subfigure}[b]{0.45\textwidth}
         \centering
         \includegraphics[width=\textwidth]{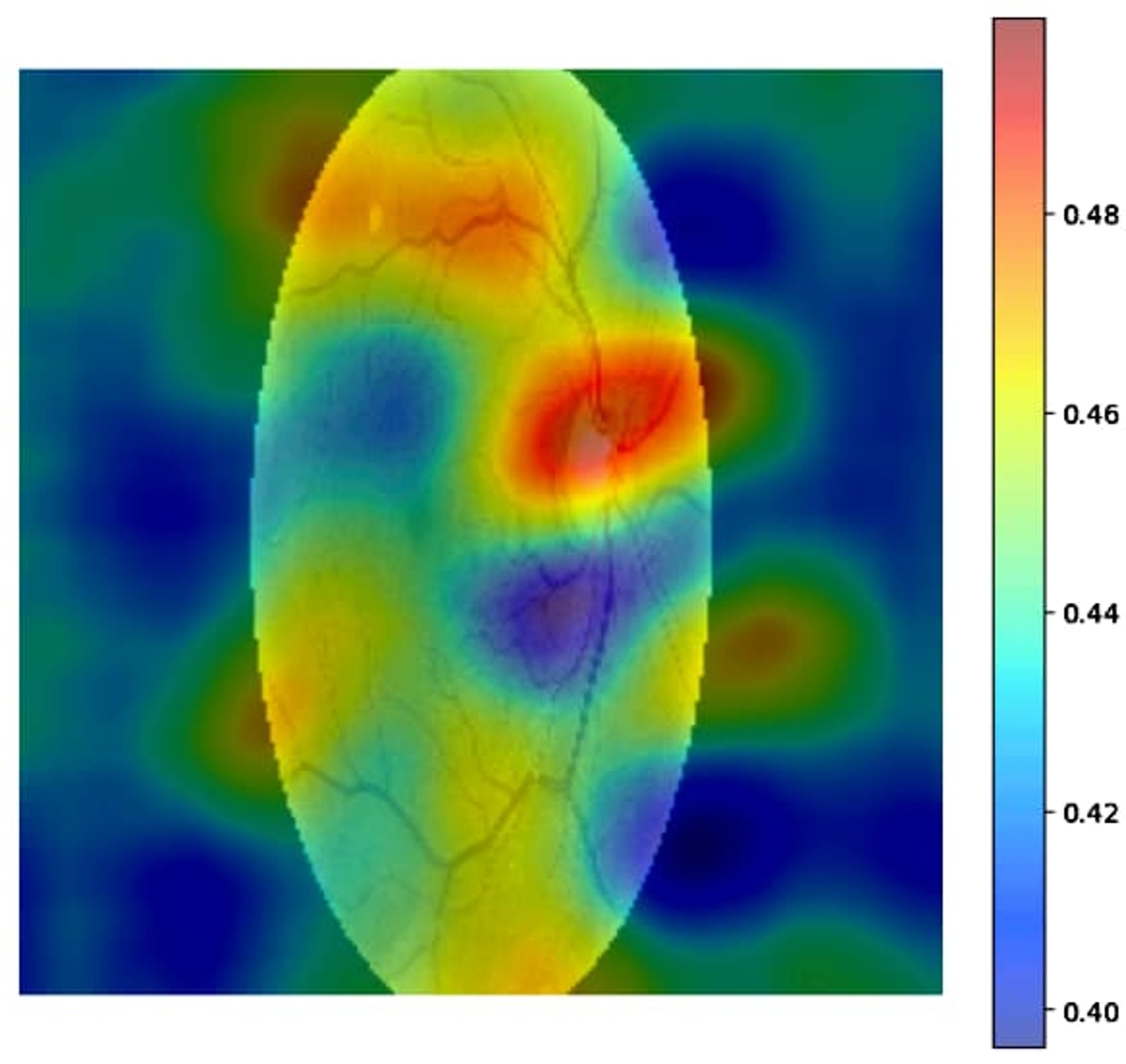}
         \caption{With dilation: Explanation for diabetes}
     \end{subfigure}
     \caption{RISE}
     \label{fig:RISE}
\end{figure}
Figure \ref{fig:RISE} illustrates the explanation by RISE 
for the input image whose label is diabetic retinopathy shown in  Figure \ref{fig:RISE}(a). 
RISE provides global insights by occluding input regions systematically.
Figure \ref{fig:RISE}(b) is the RISE explanation for other classes while Figure \ref{fig:RISE}(c) and Figure \ref{fig:RISE}(d) is the RISE explanation for without and with dilation respectively for the disease diabetic retinopathy.
These visualizations highlight the areas where the model focuses to make its decision, offering valuable insights for each class.
It can be noted that the explanation of RISE with dilation is focused on the retinal blood vessel area which is crucial for diabetes detection while it is missing in the explanation of RISE without dilation.
Additionally, the RISE explanation for other classes considered all regions except the blood vessels.
Hence, the RISE explanation shows the ResNet-152 model with dilation has achieved higher accuracy in predicting the correct classes compared to the model without dilation.

\begin{figure}[H]
     \centering
     \begin{subfigure}[b]{0.3\textwidth}
         \centering
         \includegraphics[width=\textwidth]{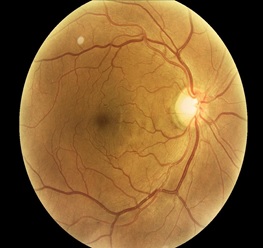}
         \caption{Input image (Diabetes)}
     \end{subfigure}
     \hfill
\begin{subfigure}[b]{0.3\textwidth}
         \centering
         \includegraphics[width=\textwidth]{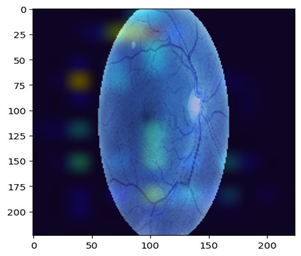}
         \caption{With dilation: Predicted as diabetes}
     \end{subfigure}
     \hfill
       \begin{subfigure}[b]{0.29\textwidth}
         \centering
         \includegraphics[width=\textwidth]
{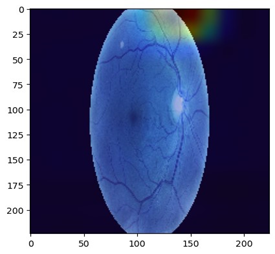}
         \caption{Without dilation: predicted as cataract}
     \end{subfigure}
     \caption{GradCAM}
     \label{fig:GradCAM}
\end{figure}
Figure \ref{fig:GradCAM} illustrates the explanation by GradCAM for the input image whose label is diabetic retinopathy shown in  Figure \ref{fig:GradCAM}(a). 
GradCAM highlights influential regions across the entire image to explain model decisions comprehensively.
Figure \ref{fig:GradCAM}(b) and Figure \ref{fig:GradCAM}(c) are the GradCAM explanation for without and with dilation respectively.
It can be noted that without dilation, the class has been incorrectly predicted as cataract, while with dilation, it has been correctly predicted as diabetes.
These visualizations highlight the areas where the model focuses to make its decision, offering valuable insights for each class.
Hence, the GradCAM explanation shows the ResNet-152 model with dilation has achieved higher accuracy in predicting the correct classes compared to the model without dilation.
The GradCAM technique further enhances our understanding by revealing the overall regions contributing to the model's decision-making across all eight disease classes.
\subsection{Analysis of Activation Maps of ResNet Models} 
The activation maps are obtained at the last convolution layer where the dilation rate has been changed. All activation maps are obtained from the trained ResNet-101 model for consistency purposes.

\begin{figure}[H]
     \centering
     \begin{subfigure}[b]{0.25\textwidth}
         \centering
         \includegraphics[width=\textwidth]{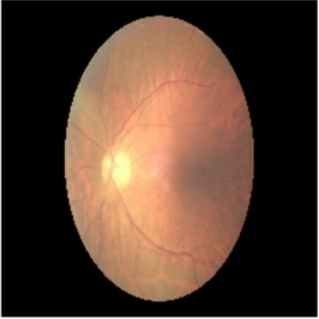}
         \caption{Cataract }
     \end{subfigure}
     \hfill
     \begin{subfigure}[b]{0.3\textwidth}
         \centering
         \includegraphics[width=\textwidth]{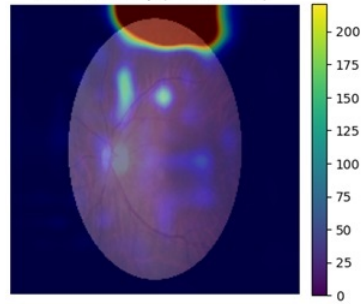}
         \caption{With dilation}
     \end{subfigure}
     \hfill
     \begin{subfigure}[b]{0.3\textwidth}
         \centering
         \includegraphics[width=\textwidth]{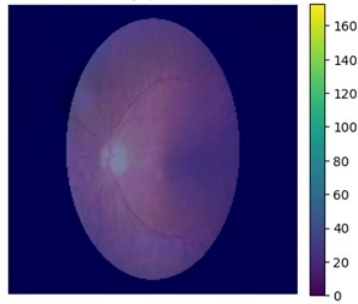}
         \caption{Without dilation}
     \end{subfigure}
     \caption{Activation map for cataract: Dilated ResNet focuses the entire region correctly}
     \label{fig:Cat}
\end{figure}

\begin{figure}[H]
     \centering
     \begin{subfigure}[b]{0.25\textwidth}
         \centering
         \includegraphics[width=\textwidth]{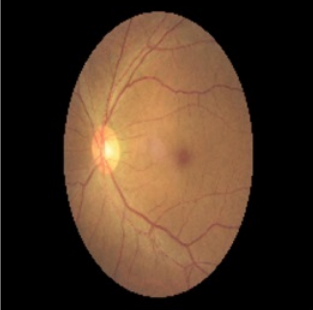}
         \caption{Normal }
            \end{subfigure}
     \hfill
     \begin{subfigure}[b]{0.3\textwidth}
         \centering         \includegraphics[width=\textwidth]{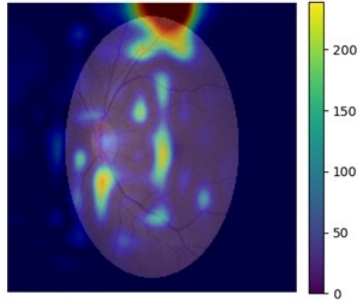}
         \caption{With dilation}
        \end{subfigure}
     \hfill
     \begin{subfigure}[b]{0.3\textwidth}
         \centering
         \includegraphics[width=\textwidth]{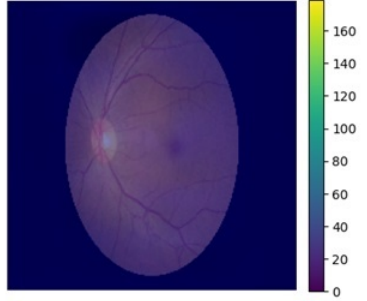}
         \caption{Without dilation}
        \end{subfigure}
     \caption{Activation map for normal: Dilated ResNet focuses the entire region correctly}
     \label{fig:Normal}
\end{figure}

\begin{figure}[H]
     \centering
     \begin{subfigure}[b]{0.25\textwidth}
         \centering
         \includegraphics[width=\textwidth]{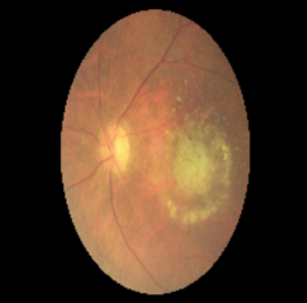}
         \caption{Abnormalities}
         \end{subfigure}
     \hfill
     \begin{subfigure}[b]{0.3\textwidth}
         \centering         \includegraphics[width=\textwidth]{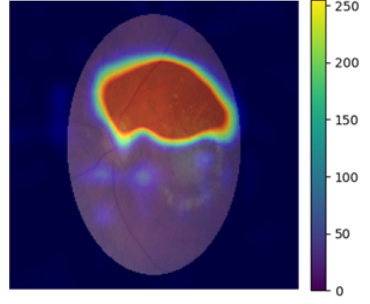}
         \caption{With dilation}
        \end{subfigure}
     \hfill
     \begin{subfigure}[b]{0.3\textwidth}
         \centering
         \includegraphics[width=\textwidth]{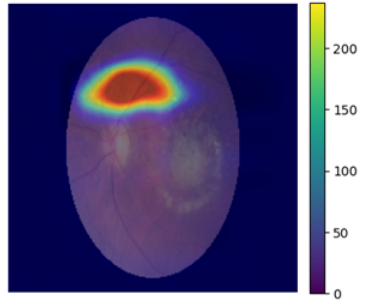}
         \caption{Without dilation}
        \end{subfigure}
     \caption{Activation map for abnormalities: Dilated ResNet focuses the entire region correctly}
     \label{fig:Abnormalities}
\end{figure}

Figure \ref{fig:Cat} shows the activation maps obtained for the class cataract with and without dilation. Similarly, Figure \ref{fig:Normal} and  Figure \ref{fig:Abnormalities} shows the activation maps obtained for the classes normal and abnormalities with and without dilation. It can be observed that the activation maps obtained with dilation correctly focus on the entire region for three classes (cataract, normal, and abnormalities), in contrast to those without dilation.

\begin{figure}[H]
     \centering
     \begin{subfigure}[b]{0.25\textwidth}
         \centering
         \includegraphics[width=\textwidth]{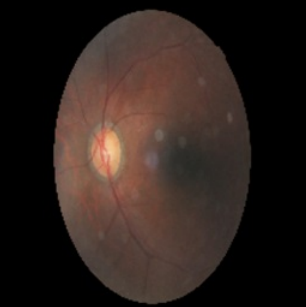}
         \caption{Glaucoma }
         \end{subfigure}
     \hfill
     \begin{subfigure}[b]{0.3\textwidth}
         \centering         \includegraphics[width=\textwidth]{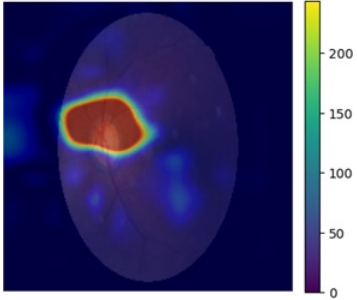}
         \caption{With dilation}
         \end{subfigure}
     \hfill
     \begin{subfigure}[b]{0.3\textwidth}
         \centering
         \includegraphics[width=\textwidth]{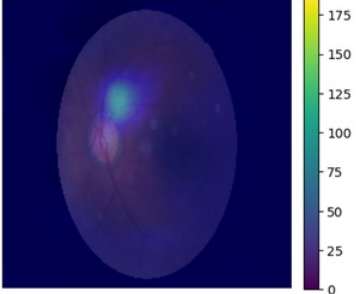}
         \caption{Without dilation}
         \end{subfigure}
     \caption{Activation map for glaucoma: Dilated ResNet focuses on the optic disc correctly}
     \label{fig:Glaucoma}
\end{figure}

\begin{figure}[H]
     \centering
     \begin{subfigure}[b]{0.25\textwidth}
         \centering
         \includegraphics[width=\textwidth]{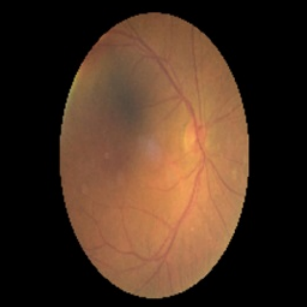}
         \caption{Degeneration }
        \end{subfigure}
     \hfill
     \begin{subfigure}[b]{0.3\textwidth}
         \centering         \includegraphics[width=\textwidth]{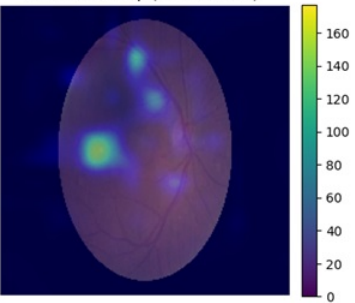}
         \caption{With dilation}
       \end{subfigure}
     \hfill
     \begin{subfigure}[b]{0.3\textwidth}
         \centering
         \includegraphics[width=\textwidth]{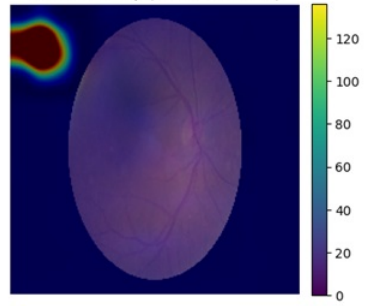}
         \caption{Without dilation}
         \end{subfigure}
     \caption{Activation map for degeneration: Dilated ResNet focuses on the macula and optic disc correctly}
     \label{fig:Degeneration}
\end{figure}

\begin{figure}[H]
     \centering
     \begin{subfigure}[b]{0.25\textwidth}
         \centering
         \includegraphics[width=\textwidth]{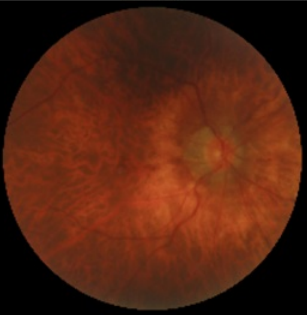}
         \caption{Myopia }
       \end{subfigure}
     \hfill
     \begin{subfigure}[b]{0.3\textwidth}
         \centering         \includegraphics[width=\textwidth]{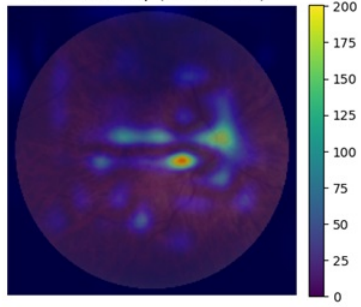}
         \caption{With dilation}
        \end{subfigure}
     \hfill
     \begin{subfigure}[b]{0.3\textwidth}
         \centering
         \includegraphics[width=\textwidth]{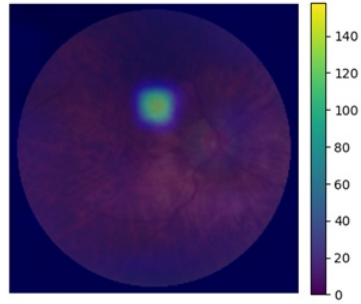}
         \caption{Without dilation}
        \end{subfigure}
     \caption{Activation map for myopia: Dilated ResNet focuses near the optic disc correctly}
     \label{fig:myo}
\end{figure}

Figure \ref{fig:Glaucoma} shows the activation maps obtained for the class glaucoma with and without dilation. 
Similarly, Figure \ref{fig:myo} and Figure \ref{fig:Degeneration} shows the activation maps
obtained for the classes myopia and degeneration with and without dilation.
It is clear from the comparison with dilation focus on the optic disc and near optic disc (macula) correctly when compared to without dilation.

\begin{figure}[H]
     \centering
     \begin{subfigure}[b]{0.25\textwidth}
         \centering
         \includegraphics[width=\textwidth]{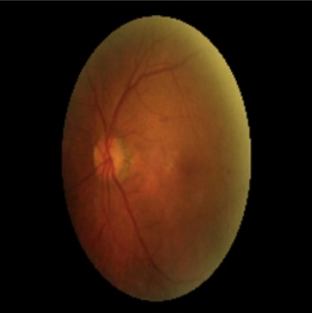}
         \caption{Diabetic}
      \end{subfigure}
     \hfill
     \begin{subfigure}[b]{0.3\textwidth}
         \centering         \includegraphics[width=\textwidth]{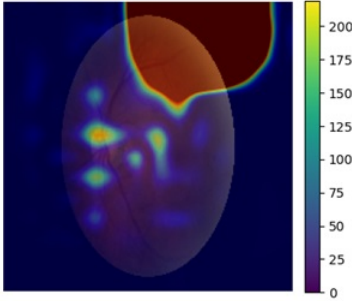}
         \caption{With dilation}
       \end{subfigure}
     \hfill
     \begin{subfigure}[b]{0.3\textwidth}
         \centering
         \includegraphics[width=\textwidth]{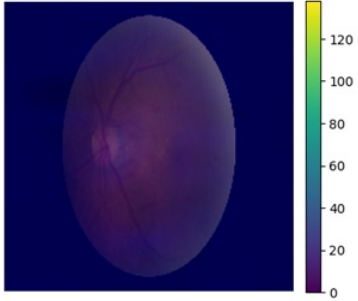}
         \caption{Without dilation}
       \end{subfigure}
     \caption{Activation map for diabetic retinopathy: Dilated ResNet focuses on the blood vessels correctly}
     \label{fig:Diabet}
\end{figure}

\begin{figure}[H]
     \centering
     \begin{subfigure}[b]{0.25\textwidth}
         \centering
         \includegraphics[width=\textwidth]{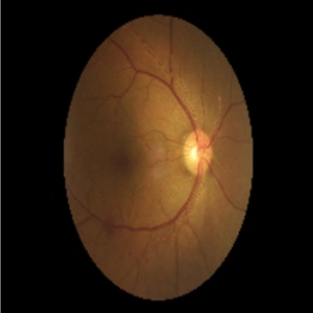}
         \caption{Hypertension}
         \end{subfigure}
     \hfill
     \begin{subfigure}[b]{0.3\textwidth}
         \centering         \includegraphics[width=\textwidth]{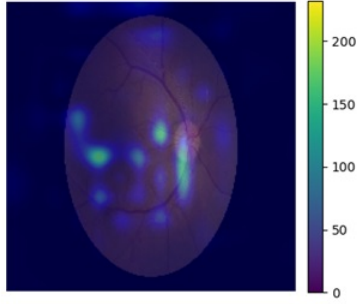}
         \caption{With dilation}
        \end{subfigure}
     \hfill
     \begin{subfigure}[b]{0.3\textwidth}
         \centering
         \includegraphics[width=\textwidth]{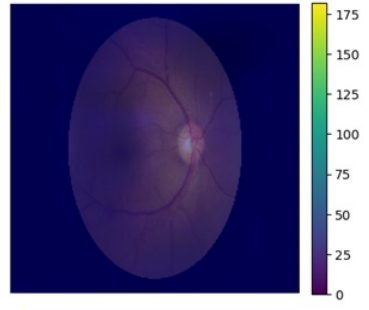}
         \caption{Without dilation}
         \end{subfigure}
     \caption{Activation map for hypertension: Dilated ResNet focuses the blood vessels correctly}
     \label{fig:hypertension}
\end{figure}

Figure \ref{fig:Diabet} shows the activation maps obtained for the class diabetic retinopathy with and without dilation. 
Similarly, Figure \ref{fig:hypertension} shows the activation maps obtained for the class hypertension with and without dilation.
It is clear from the comparison with dilation focus on the with dilation focus on blood vessels correctly when compared to without dilation.
 
From all the above activation maps, it can be observed that the ResNet model with dilation performs the best due to its focus on relevant regions for various classes. For cataract and normal conditions, the entire region is correctly focused on; for glaucoma, the focus is correctly on the optic disc; for diabetes and hypertension, the focus is correctly on the blood vessels; for degeneration, the focus is on both the macula and optic disc; for myopia, the focus is near the optic disc; and for other abnormalities, the entire region is correctly focused on.


\section{Conclusions}
\label{con}
In conclusion, this paper presents dilated Residual Network models for disease classification from retinal fundus images, demonstrating the effectiveness of dilated convolutions in improving the receptive field compared to normal convolutions in ResNet models. By incorporating computer-assisted diagnostic tools that employ deep learning enhanced with explainable AI techniques, this study aims to make the decision-making process of AI transparent, enabling medical professionals to understand and trust the AI's diagnostic decisions. This approach is particularly relevant in today's healthcare landscape, where there is a growing demand for transparency to ensure the reliability and ethical use of AI applications.

The dataset used in this work is the Ocular Disease Intelligent Recognition (ODIR), a structured ophthalmic database with eight classes covering most common retinal eye diseases. Evaluation metrics used include precision, recall, accuracy, and F1 score.
A comparative study has been conducted by applying dilation to five variants of the ResNet model: ResNet-18, ResNet-34, ResNet-50, ResNet-101, and ResNet-152. The dilated ResNet model showed promising results compared to the normal ResNet, with average F1 scores of 0.71, 0.70, 0.69, 0.67, and 0.70, respectively, for the five different versions in ODIR multiclass disease classification. These results indicate that the dilated ResNet model offers a significant improvement in feature capture and diagnostic accuracy for retinal eye diseases.

The study also shows how the dilated ResNet focuses on different regions for different diseases. For instance, it focuses on the optic disc for glaucoma, blood vessels for diabetes and hypertension, and the macula for degeneration. This disease-specific focus can be explained through XAI techniques, validating that the model’s behavior aligns with medical knowledge.

\section{Declarations}

\textbf{Conflict of Interest:} The authors declare that they have no conflict of interest.

\subsection{Data Availability}
The ODIR dataset used in the study is publicly available.\\\\
It can be downloaded from 

- https://odir2019.grand-challenge.org/ \\

\subsection{Authors' Contributions}
\begin{enumerate}
    \item P.N. Karthikayan: Literature survey, Training and testing, Verification, Drafting
    \item Yoga Sri Varshan V: Algorithm design, Conceptual design, Implementation, Analysis, Verification
    \item Hitesh Gupta Kattamuri: Training and testing, Verification, Analysis
    \item Umarani Jayaraman: Verification, Drafting, Supervision
\end{enumerate}

\bibliography{refs}

\end{document}